\documentclass[11pt]{article}
\usepackage{amsfonts,amsmath,amsxtra}
\usepackage{latexsym}
\usepackage{amssymb}

\def\hybrid{\topmargin 0pt      \oddsidemargin 0pt
        \headheight 0pt \headsep 0pt
        \textwidth 16.5cm
        \textheight 23cm
        \marginparwidth 0.0in
        \parskip 5pt plus 1pt   \jot = 1.5ex}
\catcode`\@=11
\def\marginnote#1{}
\newcount\hour
\newcount\minute
\newtoks\amorpm
\hour=\time\divide\hour by60 \minute=\time{\multiply\hour by60
\global\advance\minute by-\hour}
\edef\standardtime{{\ifnum\hour<12 \global\amorpm={am}%
        \else\global\amorpm={pm}\advance\hour by-12 \fi
        \ifnum\hour=0 \hour=12 \fi
      \number\hour:\ifnum\minute<10 0\fi\number\minute\the\amorpm}}
\edef\militarytime{\number\hour:\ifnum\minute<10 0\fi\number\minute}

\def\draftlabel#1{{\@bsphack\if@filesw {\let\thepage\relax
   \xdef\@gtempa{\write\@auxout{\string
      \newlabel{#1}{{\@currentlabel}{\thepage}}}}}\@gtempa
   \if@nobreak \ifvmode\nobreak\fi\fi\fi\@esphack}
        \gdef\@eqnlabel{#1}}
\def\@eqnlabel{}
\def\@vacuum{}
\def\draftmarginnote#1{\marginpar{\raggedright\scriptsize\tt#1}}

\def\draft{\oddsidemargin -0.1truein
        \def\@oddfoot{\sl preliminary draft \hfil
        \rm\thepage\hfil\sl\today\quad\militarytime}
        \let\@evenfoot\@oddfoot \overfullrule 3pt
        \let\label=\draftlabel
        \let\marginnote=\draftmarginnote
\def\@eqnnum{{\rm (\theequation)}
\rlap{\kern\marginparsep\tt\@eqnlabel}%
\global\let\@eqnlabel\@vacuum}  }

\newfont{\Bbbb}{msbm7 scaled 1\@ptsize00}
\newcommand{\zs}{\raise-1pt\hbox{$\mbox{\Bbbb Z}$}}

scaled 1\@ptsize00

\font\sevenmsa=msam6 
\newfam\msafam
\textfont\msafam=\sevenmsa
\def\hexnumber@#1{\ifnum#1<10 \number#1\else
\ifnum#1=10 A\else\ifnum#1=11 B\else\ifnum#1=12 C\else \ifnum#1=13
D\else\ifnum#1=14 E\else\ifnum#1=15 F\fi\fi\fi\fi\fi\fi\fi}
\def\msa@{\hexnumber@\msafam}
\def\llcorner{\delimiter"4\msa@78\msa@78 }
\def\lrcorner{\delimiter"5\msa@79\msa@79 }
\mathchardef\blacktriangleright="3\msa@49
\mathchardef\blacktriangleleft="3\msa@4A \font\tenmsb=msbm10 scaled
1\@ptsize00
\newfam\msbfam
\textfont\msbfam=\tenmsb \scriptfont\msbfam=\tenmsb

\newdimen\Squaresize \Squaresize=14pt
\newdimen\Thickness \Thickness=0.5pt

\def\Square#1{\hbox{\vrule width \Thickness
   \vbox to \Squaresize{\hrule height \Thickness\vss
      \hbox to \Squaresize{\hss#1\hss}
   \vss\hrule height\Thickness}
\unskip\vrule width \Thickness} \kern-\Thickness}

\def\Vsquare#1{\vbox{\Square{$#1$}}\kern-\Thickness}

\def\numberbysection{\@addtoreset{equation}{section}
        \def\theequation{\thesection.\arabic{equation}}}
\numberbysection

\renewcommand{\theequation}{\thesection.\arabic{equation}}
\def\titlepage{\@restonecolfalse\if@twocolumn\@restonecoltrue\onecolumn
     \else \newpage \fi \thispagestyle{empty}\c@page\z@
        \def\thefootnote{\fnsymbol{footnote}} }

\def\endtitlepage{\if@restonecol\twocolumn \else  \fi
        \def\thefootnote{\arabic{footnote}}
        \setcounter{footnote}{0}}  
\relax

\hybrid
\makeatletter
\newdimen\normalarrayskip            
\newdimen\minarrayskip               
\normalarrayskip\baselineskip \minarrayskip\jot
\newif\ifold             \oldtrue            \def\new{\oldfalse}
\def\arraymode{\ifold\relax\else\displaystyle\fi}
\def\eqnumphantom{\phantom{(\theequation)}} 
\def\@arrayskip{\ifold\baselineskip\z@\lineskip\z@
     \else
     \baselineskip\minarrayskip\lineskip1\baselineskip\fi}

\def\@arrayclassz{\ifcase \@lastchclass \@acolampacol \or
\@ampacol \or \or \or \@addamp \or
   \@acolampacol \or \@firstampfalse \@acol \fi
\edef\@preamble{\@preamble
  \ifcase \@chnum
     \hfil$\relax\arraymode\@sharp$\hfil
     \or $\relax\arraymode\@sharp$\hfil
     \or \hfil$\relax\arraymode\@sharp$\fi}}

\def\@array[#1]#2{\setbox\@arstrutbox=\hbox{\vrule
     height\arraystretch \ht\strutbox
     depth\arraystretch \dp\strutbox
width\z@}\@mkpream{#2}\edef\@preamble{\halign \noexpand\@halignto
\bgroup \tabskip\z@ \@arstrut \@preamble \tabskip\z@ \cr}%
\let\@startpbox\@@startpbox \let\@endpbox\@@endpbox
  \if #1t\vtop \else \if#1b\vbox \else \vcenter \fi\fi
  \bgroup \let\par\relax
  \let\@sharp##\let\protect\relax
  \@arrayskip\@preamble}
\def\eqnarray{\stepcounter{equation}%
              \let\@currentlabel=\theequation
              \global\@eqnswtrue
              \global\@eqcnt\z@
              \tabskip\@centering              
              \let\\=\@eqncr
              $$%
            \halign to \displaywidth  \bgroup
             \eqnumphantom \@eqnsel
      \hskip\@centering                               
    $\displaystyle  \tabskip\z@ {##}$%
    &\global\@eqcnt\@ne \hskip 2\arraycolsep
         $ \displaystyle  \arraymode{##}$\hfil
    &\global\@eqcnt\tw@ \hskip 2\arraycolsep
         $\displaystyle\tabskip\z@{##}$\hfil
         \tabskip\@centering
    &{##}\tabskip\z@\cr}
\makeatother

\newcommand{\CC}{{\mathbb{C}}}

\def\IC{\mathbb{C}}

\def\IP{\mathbb{P}}

\def\IR{\mathbb{R}}
\def\IZ{\mathbb{Z}}
\def\CA {\mathcal{A}}

\def\CC {\mathcal{C}}

\def\CE {\mathcal{E}}

\def\CG {\mathcal{G}}

\def\CL {\mathcal{L}}
\def\CM {\mathcal{M}}
\def\CN {\mathcal{N}}
\def\CO {\mathcal{O}}

\def\CV {\mathcal{V}}

\def\la{\lambda}

\def\pr {\partial}
\def\apr {\overline {\partial }}
\def\ib{\bar{i}}
\def\jb{\bar{j}}
\def\kb{\bar{k}}
\def\lb{\bar{l}}

\def\wb {\bar{w}}
\def\zb {\bar{z}}

\def\nn{\nonumber}

\newtheorem{te}{Theorem}[section]

\newtheorem{prop}{Proposition}[section]
\newtheorem{cor}{Corollary}[section]
\newtheorem{lem}{Lemma}[section]

\newtheorem{rem}{Remark}[section]

\newcommand{\proof}{\noindent {\it Proof}.\,\,}
\newcommand\bqa{\begin{eqnarray}}
\newcommand\eqa{\end{eqnarray}}
\def\be{\begin{eqnarray}\new\begin{array}{cc}}
\def\ee{\end{array}\end{eqnarray}}
\def\nn{\nonumber}

\def\beq{\begin{equation}}
\def\eeq{\end{equation}}
\def\bse{\begin{subequations}}                
\def\ese{\end{subequations}}
\def\bp{\begin{pmatrix}}
\def\ep{\end{pmatrix}}

\def\h{\hbar}
\def\i{\imath}

\def\stack#1#2{\raise0.7pt\hbox{$\mathrel{\mathop{#2}\limits^{#1}}$}}
\def\tr{\triangleright}
\def\tl{\triangleleft}
\def\sem{\mathsurround=0pt \raise1pt
\hbox{$\scriptscriptstyle>\!\!$}\:\!\!\tl}
\def\mes{\mathsurround=0pt \tr\!\:\!\raise0.8pt
\hbox{$\scriptscriptstyle\!\!<$}\,}
\def\]{\mathsurround=0pt ]\raise-2pt\hbox{$_\ast$}}

\def\<{\langle}
\def\>{\rangle}

\def\CO{{\cal O}}

\def\we{\raise-1pt\hbox{$\,\stackrel{\wedge}{,}\,$}}
\def\tr{{\rm tr}\,}

\def\pr {\partial}

\def\vp{\varphi}
\newcounter{pac}[section]

\newcounter{pacc}[subsection]

0

\title{\bf Archimedean $L$-factors and \\
Topological Field Theories II}
\begin{document}
\author{Anton Gerasimov, Dimitri Lebedev and Sergey Oblezin}
\date{}
\maketitle

\renewcommand{\abstractname}{}

\begin{abstract}
\noindent {\bf Abstract}.  In the first part of this series  of
papers we  propose  a functional integral representation for local
Archimedean $L$-factors given by  products  of the
$\Gamma$-functions. In particular we derive  a representation of the
$\Gamma$-function as a properly regularized equivariant symplectic
volume of an infinite-dimensional space. The corresponding
functional integral arises in the description of a type $A$
equivariant topological linear sigma model on a disk. In this paper
we provide a functional integral representation of the Archimedean
$L$-factors in terms of a type $B$ topological sigma model on a
disk. This representation leads naturally to the classical
Euler integral representation of the $\Gamma$-functions. These two
integral representations of $L$-factors in terms of $A$ and $B$
topological sigma models are related by a mirror map. The mirror
symmetry in our setting should be considered as a local Archimedean
Langlands correspondence between two constructions of local
Archimedean $L$-factors.

\end{abstract}
\vspace{5 mm}

\section*{Introduction}

In \cite{GLO1} we propose a framework of topological quantum field
theory as a proper  way to describe arithmetic  geometry of
Archimedean places of the compactified spectrum of global number
fields. In particular we provide a functional integral
representation of local Archimedean $L$-factors as correlation
functions in  two-dimensional type $A$ equivariant topological sigma
models.  This representation implies that local Archimedean
$L$-factors are equal to  properly defined equivariant symplectic
volumes of  spaces of holomorphic maps of a  disk into complex
vector spaces. Thus the equivariant infinite-dimensional symplectic
geometry (in the framework of a topological quantum
field theory) appears as the Archimedean counterpart  of the
geometry over non-Archimedean local fields.

The construction of local Archimedean $L$-factors in terms of type
$A$ equivariant topological sigma models should be considered as an
analog of ``arithmetic'' construction of local non-Archimedean
$L$-factors in terms of representations of local non-Archimedean
Galois group. There is another, ``automorphic'' construction of the
non-Archimedean $L$-factors which uses  a theory of
infinite-dimensional representations of reductive groups. For
Archimedean places this provides a representation of the
corresponding $L$-factors as  products of  classical Euler's
integral representations of the $\Gamma$-functions.   In \cite{GLO1}
we conjecture  that this finite-dimensional integral representation
of $L$-factors naturally arises in a type $B$ topological sigma
model which is mirror dual to the type $A$ topological sigma model
considered in \cite{GLO1}. This would lead to an identification  of
local Archimedean Langlands correspondence between ``arithmetic''
and ``automorphic'' constructions of $L$-functions with a mirror
symmetry  between corresponding type $A$ and type $B$ equivariant
topological sigma models.  In this note we propose the type $B$
topological sigma model dual to the one considered in \cite{GLO1}
and identify a particular set of correlation functions on a disk
with Archimedean $L$-functions. As expected the resulting functional
integral representation  of the $L$-factors is reduced to a product
of the  Euler integral  representations of  $\Gamma$-functions.

The type $B$  equivariant topological sigma model considered below
is  an $S^1$-equivariant sigma model on a disk $D$ with the target space
$X=\IC^{\ell+1}$ and a non-trivial superpotential $W$. We imply that
$S^1$ acts by rotations of the disk $D$. A particular superpotential $W$
corresponding to the mirror dual to the type $A$ equivariant
topological sigma model with target space $\IC^{\ell+1}$ is
well-known \cite{HV}.  However our considerations have some new
interesting features. At first, the $S^1$-equivariance
provides a new solution of the so-called Warner problem in
topological theories on non-compact manifolds. The standard way to
render the theory consistent is to introduce a non-trivial boundary
interaction corresponding to a collection of $D$-branes in the
target space \cite{K}, \cite{KL}, \cite{O}. We show that in the case
of $S^1$-equivariant sigma model on the disk $D$ there is a
universal boundary term leading to a consistent topological theory.
Another not quite standard feature of our approach  is a choice of a
real structure on the space of fields of the topological theory. One
can  construct  a topological quantum field theory starting  with an $\CN=2$ SUSY
quantum field theory and using  a twisting procedure (see e.g. \cite{W1},
\cite{W2}).  This provides a particular real structure on the space of
fields. Another approach is to construct directly topological
theory combining (equivariant) topological multiplets of quantum
fields. Although this approach
produces topological
 field theories closely related with  those obtained by the twisting
 procedure the resulting  real structure  may be different
(for a discussion of an example see e.g.  \cite{W1}). In our
considerations we use a real structure which is different from the
one appeared in twisted  $\CN=2$ SUSY
two-dimensional sigma models.

We also comment on an  explicit mirror map of type $A$ and type $B$
topological sigma models.  We provide a heuristic derivation of the
$B$-model superpotential $W$ by applying Duistermaat-Heckman
localization formula to an infinite-dimensional projective space.
The sum over fixed points can be related  to the sum over
instantons used in the previous derivations of the superpotential
\cite{HV}. We also consider an explicit change of variables in the
functional integral transforming $A$-model into $B$-model. Although
these considerations are heuristic they reveal  interesting
features of the topological theories discussed in this note and in
\cite{GLO1}.

The plan of the paper is as follows. In Section 1 we provide  a
 construction of a $S^1$-equivariant type $B$ topological sigma model
on a disk $D$.  In Section 2 we
identify a particular correlation function of the topological sigma
model with a product of $\Gamma$-functions thus providing a new
functional integral representation of local Archimedean $L$-factors.
In Section 3 we give  heuristic constructions
of a mirror map of type $A$ topological sigma model considered in
\cite{GLO1} to a type $B$ topological sigma model considered in
Section 2. In Section 4 we conclude with some general remarks
and discuss further directions of research.

{\em Acknowledgments}: The research was  partly supported by
Grants RFBR-08-01-00931-a, 09-01-93108-NCNIL-a.  The research of
SO was partly supported by RF President grant MK-544.2009.1.
 AG was  also partly supported by Science Foundation Ireland grant.

\section{Type $B$ Topological sigma-models}

We start by recalling the standard construction of a topological sigma
model  associated with a K\"{a}hler manifold
with trivial canonical class supplied with holomorphic
superpotential.
For general discussion of the two-dimensional topological
sigma models see e.g.   \cite{CMR} and reference therein.

Let $X$ be a  K\"{a}hler manifold of complex dimension $(\ell+1)$
with trivial canonical class and let $W\in H^0(X,\CO)$. Let
 $\CM(\Sigma,X)=Map(\Sigma,X)$ be the space of maps $\Phi:\,\Sigma\to X$ of a
compact Riemann surface $\Sigma$ into $X$. Let $(z,\zb)$ be local
complex coordinates on  $\Sigma$. We pick a hermitian metric $h$ on
$\Sigma$ and denote $\sqrt{h}\,d^2z$ the corresponding measure on
$\Sigma$. The complex structure on $\Sigma$ defines a decomposition
$d=\pr+\apr$, $\pr=dz\,\pr_z$, $\apr=d\zb\,\pr_{\zb}$ of the
differential $d$ acting on the differential forms on $\Sigma$. Let
$K$ and $\bar{K}$ be canonical and anti-canonical bundles over
$\Sigma$. Let $\omega$ and $g$ be the K\"{a}hler form   and the
K\"{a}hler metric on $X$ and  $T_{\IC}X=T^{1,0}X\oplus T^{0,1}X$ be
a decomposition of the complexified tangent bundle of $X$. We choose
local complex coordinates $(\phi^j$,$\bar{\phi}^{j})$  on $X$.
Locally Levi-Civita  connection $\Gamma$ and the corresponding
Riemann  curvature tensor  $R$ are  given by
 \be
  \Gamma_{jk}^i=g^{i\bar{n}}\pr_{j}g_{k\bar{n}},\qquad
  R_{i\bar{j}k\bar{l}}=g_{m\bar{j}}\pr_{\bar{l}}\Gamma^m_{ik}.
 \ee
Now let us specify the standard field content of the type $B$
topological sigma model associated with a pair $(X,W)$. Denote $\Pi$
the parity change functor. Thus $\Pi\CE$ is a bundle $\CE$ with the
opposite parity of the fibers.
 Let  $\eta$, $\theta$ be sections of $\Phi^*(\Pi T^{0,1}X)$,
$\rho$ be a section of $(K\oplus \bar{K})\otimes \Phi^*(\Pi T^{1,0}X)$.
We also introduce the fields $\bar{G}$ and $G$
given by sections of $\Phi^*(T^{0,1}X)$ and $K\otimes \bar{K}\otimes
\Phi^*(T^{1,0}X)$ respectively.  The  BRST transformation $\delta$ is defined as
follows:
\be\label{BRST}
\delta \bar{\phi}^{\ib}=\bar{\eta}^{\ib},\qquad
\delta \bar{\eta}^{\ib}=0,\qquad
\delta\theta^{\ib}=\bar{G}^{\ib}-\Gamma^{\ib}_{\jb
  \kb}\bar{\eta}^{\jb}
\theta^{\kb}, \qquad \delta \bar{G}^{\ib}=-\Gamma^{\ib}_{\jb
\kb}\bar{G}^{\jb}\bar{\eta}^{\kb}, \ee \be\nn \delta
\rho^i=-d\phi^i, \qquad \delta \phi^i=0, \qquad  \delta
G^i=d\rho^{i}+\Gamma_{jk}^id\phi^j\wedge \rho^k+ \frac{1}{2}R^i_{j
k\bar{l}}\bar{\eta}^{\bar{l}} \rho^j\wedge
 \rho^k.
 \ee
Straightforward calculations show that $\delta^2=0$. One can define
new variables \be\label{newvar}
\bar{\CG}^{\ib}=\bar{G}^{\ib}-\Gamma_{\jb\kb}^{\ib}\bar{\eta}^{\jb}\theta^{\kb},\qquad
\CG^{i}=G^{i}+\frac{1}{2}\Gamma_{j k}^{i}\rho^{j}\wedge \rho^{k},
\ee such that the action of $\delta$ has the following  canonical
form: \be\label{modtrnas} \delta
\bar{\phi}^{\ib}=\bar{\eta}^{\ib},\qquad \delta
\bar{\eta}^{\ib}=0,\qquad \delta\theta^{\ib}=\bar{\CG}^{\ib}, \qquad
\delta \bar{\CG}^{\ib}=0, \ee \be\nn \delta \rho^i=-d\phi^i, \qquad
\delta \phi^i=0, \qquad  \delta \CG^i=d\rho^{i}. \ee Here the
property $\delta^2=0$ is obvious. The advantage of \eqref{BRST} is
that the fields $G^i$ and  $\bar{G}^{\jb}$ are covariant with
respect to  diffeomorphisms of the target space $X$.

Consider a  topological sigma model with the action given by
\be\label{totact}
S=S_0+S_{\bar{W}}+S_{W},
\ee
where
\be\label{S1}
S_0=\int_{\Sigma}\,(g_{i\jb} d
\phi^i\wedge*d\bar{\phi}^{\jb}+g_{i\jb}\rho^i\wedge *D\bar{\eta}^{\jb}
-\,g_{i\jb}\theta^{\jb}D\rho^i+g_{i\jb}
G^i\bar{G}^{\jb}-\frac{1}{2} R_{i\bar{l}k\bar{j}}\bar{\eta}^{\bar{l}}\,\theta^{\jb}\,
\rho^i \wedge \rho^{k}),
\ee
\be\label{S2}
S_{\bar{W}}=\int_{\Sigma}d^2z
\sqrt{h}\left(D_{\ib}\pr_{\jb}\bar{W}(\bar{\phi})\,\,
\bar{\eta}^{\ib}\theta^{\jb}+
\bar{G}^{\ib}\,\pr_{\ib}\bar{W}(\bar{\phi})\right),
\ee
\be\label{S3}
S_{W}=\int_{\Sigma}\,\left(-\frac{1}{2}
 D_i\pr_jW(\phi)\,\rho^i\wedge \rho^j+G^i\,\pr_iW(\phi)\right),
\ee
and
$$
D_i\pr_jW(\phi)=\pr_i\pr_jW-\Gamma_{ij}^k\pr_kW,\qquad
D\bar{\eta}^{\jb}=d\bar{\eta}^{\jb}+\Gamma_{\kb\bar{\ell}}^{\jb}d\phi^{\kb}\eta^{\bar{\ell}}.
$$
The Hodge $*$-operator acts on one forms as follows  $*dz=\imath dz$,
$*d\zb=-\imath d\zb$.

The parts $S_0$ and $S_{\bar{W}}$ are  $\delta$-exact
as it follows from $\delta^2=0$ and the following representation
$$
S_0=\int_{\Sigma}\,\delta \CV_0,\qquad
S_{\bar{W}}=\int_{\Sigma}d^2z\,\sqrt{h}\,\delta \CV_{\bar{W}},
$$
where
\be\label{Fermion}
\CV_0=-g_{i\jb}\rho^i\wedge *d\bar{\phi}^{\jb}+G^i\theta_i,
\qquad \CV_{\bar{W}}=\,\theta^{\jb}\pr_{\jb}\bar{W}(\bar{\phi}),
\ee
and  $\theta_i=g_{i\bar{j}}\theta^{\bar{j}}$. The variation of $S_W$ is given by
\be\label{warner}
\delta S_W=\int_{\Sigma}\,d(\rho^i\pr_iW(\phi)),
\ee
and thus is trivial on a compact surface $\Sigma$.
Note that the  action $S_W$ is $\delta$-closed but does not $\delta$-exact.

In this paper we consider a particular case
of an equivariant  type B topological sigma model on a non-compact
two-dimensional manifold $\Sigma$.
Let $\Sigma$ be a disk $D=\{z\in \IC|\,\,|z|\leq 1\}$. We fix a
flat metric $h$  on $D$ \be\label{metricD} h=\frac{1}{2}(dz
d\zb+d\zb\,dz)= (dr)^2+r^2(d\sigma)^2, \qquad r\in [0,1],\quad
\sigma\in [0,2\pi], \ee where $z=re^{\imath \sigma}$. This metric is
obviously invariant with respect to the rotation group $S^1$
acting by $\sigma\to \sigma+\alpha$.

We would like to consider
an $S^1$-equivariant version of the type $B$  topological linear sigma model on a disk
$D$ with a superpotential $W$.  To construct an  $S^1$-equivariant extension
of the topological field theory
we modify the $\delta$-transformations  taking into account
an interpretation of $\delta$ as the de Rham differential in the infinite-dimensional
setting. Let us first recall a construction of an
algebraic model of $S^1$-equivariant cohomology.
Let $M$ be a $2(\ell+1)$-dimensional manifold supplied with
an action of  $S^1$. The Cartan  algebraic model of  $S^1$-equivariant de Rham
cohomology $H^*_{S^1}(M)$ is  the following equivariant
extension $(\Omega_{S^1}^*(M),d_{S^1})$ of
the standard de Rham complex $(\Omega^*(M),d)$:
\be\label{equivdif}
\Omega^*_{S^1}(M)=(\Omega^*(M))^{S^1}\otimes \IC[\hbar], \qquad
d_{S^1}=d+\hbar \iota_{v_0},
\ee
where $(\Omega^*(M))^{S^1}$ is an $S^1$-invariant part of
$\Omega^*(M)$, $\hbar$ is a generator of the ring $H^*(BS^1)$ and $v_0$ is a
vector field on $M$ corresponding to a generator of ${\rm Lie}(S^1)$.  We have
\be
d_{S^1}^2=\h\CL_{v_0},\qquad \CL_{v_0}=d\,\iota_{v_0}+\iota_{v_0}\,d,
\ee
where $\CL_{v_0}$ is  the Lie derivative along the vector field $v_0$.
The equivariant differential $d_{S^1}$ satisfies  $d^2_{S^1}=0$
 when  acting  on $\Omega^*_{S^1}(M)$.   The cohomology groups
$H^*_{S^1}(M)$ of the complex \eqref{equivdif}  have a natural
structure of modules over $H^*_{S^1}({\rm pt})=\IC[\hbar]$.

The $S^1$-equivariant version of the BRST transformations
\eqref{modtrnas} is a direct generalization of the expression
\eqref{equivdif} for the equivariant differential
 to the infinite-dimensional setting. Taking into account an
induced action of $S^1$ on the space of fields we have
$$
\delta_{S^1}\bar{\phi}^{\ib}=\bar{\eta}^{\ib},\qquad
\delta_{S^1}\bar{\eta}^{\ib}=\hbar\iota_{v_0}d\bar{\phi}^{\ib},\qquad
\delta_{S^1}\theta^{\ib}=\bar{\CG}^{\ib},\qquad
\delta_{S^1}\bar{\CG}^{\ib}=\hbar\iota_{v_0}d\theta^{\ib},
$$
$$
\delta_{S^1}\CG^{i}=d\rho^i,\qquad
\delta_{S^1}\rho^i=-d\phi^i-\hbar\iota_{v_0}\CG^i,\qquad
\delta_{S^1}\phi^i=\hbar \iota_{v_0}\rho^i.
$$
Obviously we have $\delta_{S^1}^2=\h\CL_{v_0}$.

In terms of the variables $G^i$ and $\bar{G}^i$ we have the following
 transformations:
$$
 \begin{array}{c}
  \delta_{S^1}\bar{\phi}^{\ib}\,=\,\bar{\eta}^{\ib}\,,
  \hspace{2cm}
  \delta_{S^1}\bar{\eta}^{\ib}\,=\,\hbar\iota_{v_0}d\bar{\phi}^{\ib}\,,
  \hspace{2cm}
  \delta_{S^1}\theta^{\ib}\,=\,\bar{G}^{\ib}
  \,-\,\Gamma^{\ib}_{\jb\kb}\bar{\eta}^{\jb}\theta^{\kb}\,,\\
 \\
  \delta_{S^1}\bar{G}^{\ib}\,=\,-\Gamma^{\ib}_{\jb\kb}
\bar{\eta}^{\jb}\bar{G}^{\kb}+\h\iota_{v_0}\bigl(
  D\theta^{\ib})
  +\hbar\pr_l\Gamma^{\ib}_{\jb\kb}(\iota_{v_0}\rho^l)\bar{\eta}^{\jb}\theta^{\kb}\,,\\
 \\
  \delta_{S^1}G^i\,=\,d\rho^i+\Gamma^i_{jk}d\phi^j\wedge\rho^k
  +\frac{1}{2}R^i_{jk\lb}\bar{\eta}^{\lb}\rho^j\wedge\rho^k
  +\h\Gamma^i_{jk}(\iota_{v_0}G^j)\wedge\rho^k\,,\\
 \\
  \delta_{S^1}\rho^i\,=\,-d\phi^i-\hbar\iota_{v_0}G^i
  -\hbar\Gamma^i_{jk}(\iota_{v_0}\rho^j)\rho^k\,,
  \hspace{2.5cm}
  \delta_{S^1}\phi^i\,=\,\hbar\iota_{v_0}\rho^i\,.
 \end{array}
$$
Now the $S^1$-equivariant version of \eqref{S1} and \eqref{S2}
on a disk $\Sigma=D$
is obtained by applying modified $\delta_{S^1}$ to $\CV_0$ and $\CV_{\bar{W}}$
given by \eqref{Fermion}. The action $S_W$ given by \eqref{S3}
 is not $\delta_{S^1}$
invariant on the disk and needs a correction boundary term.

\begin{prop}
The following modified action functional of a type $B$ topological sigma
model
\be\nn
S=\int_{D}\,\left(g_{i\jb}\left(d\phi^j+\hbar \iota_{v_0}G^j\right)
\wedge *d\bar{\phi}^{\jb}+g_{i\jb}\rho^i\wedge *D\bar{\eta}^{\jb}
-g_{i\jb}\theta^{\jb}D\rho^i+g_{i\jb}G^i\bar{G}^{\jb}-
\frac{1}{2} R_{i\bar{l}k\bar{j}}\bar{\eta}^{\bar{l}}\,\theta^{\jb}\,
\rho^i \wedge \rho^{k}\right)
\ee
\be\nn
+\int_{D}d^2z
\sqrt{h}\left(D_{\ib}\pr_{\jb}\bar{W}(\bar{\phi})
\bar{\eta}^{\ib}\theta^{\jb}+
\pr_{\ib}\bar{W}(\bar{\phi})\,\bar{G}^{\ib}\right)+\int_{D}\,\,\left(
 -\frac{1}{2} D_i\pr_jW(\phi)\,\rho^i\wedge\rho^j+\pr_iW(\phi)\,G^i\right)
\ee
\be\label{totalboundary}
-
\frac{1}{\hbar}\,\int_{S^1=\pr D}d\sigma \,\,\,W(\phi)
\ee
is $\delta_{S^1}$-invariant.
\end{prop}

\proof Direct calculation shows that $\delta_{S^1}$-variation
of the sum of the integrals over $D$ in \eqref{totalboundary}
is given by the boundary term
$$
\delta_{S^1} S=\int_{\pr D}\,\rho^i\pr_iW(\phi).
$$
The  $\delta_{S^1}$-variation of the boundary term
in \eqref{totalboundary} precisely cancels this contribution. $\Box$

\begin{rem} The action \eqref{totalboundary} does not have a smooth
  limit $\hbar\to 0$. This is a so called ``Warner
  problem'' in  the type $B$ topological sigma model
with a non-trivial superpotential $W\in H^0(X,\CO)$ on   non-compact surface
$\Sigma$. In non-equivariant setting it is resolved by
imposing special boundary conditions corresponding to
a collection of $D$-branes on the target space $X$ \cite{K},
\cite{KL}, \cite{O}. Remarkably the $S^1$-equivariant setting
discussed above allows a construction of a universal $\delta_{S^1}$-invariant boundary
condition by adding boundary term \eqref{totalboundary}.
\end{rem}

\begin{rem} The relation between the boundary term in \eqref{totalboundary}
and the variation \eqref{warner} is a particular instance of a general
descent relation between various observables in topological field
theories. 
\end{rem}

\section{Linear sigma model  on a disk}

In this Section we calculate a particular correlation function of
the $S^1$-equivariant  type $B$ linear
sigma model on the disk $D$ with the target space $\IC^{\ell+1}$ and
a generic  superpotential $W$.
The $\delta_{S^1}$-transformations in the case of $X=\IC^{\ell+1}$
are given by
\be\label{linearS11}
\delta_{S^1} \bar{\phi}^{\ib}= \bar{\eta}^{\ib},\qquad
\delta_{S^1} \bar{\eta}^{\ib}=\hbar \iota_{v_0}d\bar{\phi}^{\ib},
\qquad \delta_{S^1}\theta^{\ib}=\bar{G}^{\ib},\qquad
\delta_{S^1} \bar{G}^{\ib}=\hbar \iota_{v_0}d\theta^{\ib},
\ee
$$
\delta_{S^1} \rho^i=-d\phi^i-\hbar \iota_{v_0}G^i, \qquad \delta_{S^1}
 \phi^i=\hbar \iota_{v_0}\rho^i,
 \qquad \delta_{S^1} G^i=d\,\rho^{i}.
$$
The action \eqref{totalboundary} in this case is reduced to
\be\label{linearTFT}
S=\sum_{j=1}^{\ell+1}\int_{D}\, \left(
(d\phi^j+\hbar \iota_{v_0}G^j)\wedge *d\bar{\phi}^{j}+\rho^j\wedge
*d\bar{\eta}^{j}-\theta_jd\rho^j+G^j\bar{G}^{j}\right)
\ee
$$
+\sum_{i,j=1}^{\ell+1}
\int_{D}d^2z
\sqrt{h}\left(\apr_{i}\apr_{j}\bar{W}(\bar{\phi})\bar{\eta}^{i}\bar{\theta}^{j}+
  \apr_{i}\bar{W}(\bar{\phi})\,\bar{G}^{i}\right)+\int_{D}\,\left(-\frac{1}{2}
  \pr_i\pr_jW\,\rho^i\wedge \rho^j+\pr_iW\,G^i\right)
$$
$$
-\frac{1}{\hbar}\int_{S^1=\pr D}\,\,d\sigma W(\phi).
$$
Topological linear sigma model \eqref{linearTFT} allows a
non-standard real structure. This means the following. Let us
consider the fields $\phi^i$, $\bar{\phi}^i$, $\theta_i$,
$\bar{\theta}_i$,  $\bar{\eta}^i$, $\eta^i$, $\rho^i$
$\bar{\rho}^i$, $G^i$, $\bar{G}^i$ as independent complex fields.
The subspace of the fields entering  the description of the
topological theory with the action \eqref{linearTFT} is defined as a
subspace invariant with respect to an involution  acting as follows:
\be\label{oldRS} (\phi^i)^\dagger=\bar{\phi}^i,\qquad
(\theta_i)^\dagger=\bar{\theta}_i,\qquad
(\bar{\eta}^i)^\dagger=\eta^i, \qquad
(\rho^i)^\dagger=\bar{\rho}^i,\qquad (G^i)^\dagger=\bar{G}^i. \ee
The involution defines a real structure on the space of fields. One
can however consider another real structure defined by the reality
conditions \be\label{newRS} (\phi^i)^\dagger=\phi^i,\qquad
(\bar{\phi}^i)^\dagger=-\bar{\phi}^i,\qquad
(\theta_i)^\dagger=-\theta_i, \ee \be\nn \qquad
(\bar{\eta}^i)^\dagger=-\bar{\eta}^i, \qquad
(\rho^i)^\dagger=\rho^i,\qquad (G^i)^\dagger=G^i,\qquad
(\bar{G}^i)^\dagger=-\bar{G}^i. \ee Thus for example the fields
$\phi^i$ and $\imath \bar{\phi}^i$ are real independent fields.  To
distinguish the real fields in the sense \eqref{newRS} let us
introduce new notations  $\phi_+^i$, $\phi_-^{i}$, $G^i_+$, $G^i_-$
for $\phi^i$, $\imath \bar{\phi}^{i}$, $G^i$, $\imath \bar{G}^i$.
Similarly we redefine  the fields $\bar{\eta}$ and $\theta$ by
multiplying them on $\imath$ and considering the resulting fields as
real ones. The $S^1$-equivariant BRST operator can be defined on the
new set of real fields as follows:
 \be\label{linearS1} \delta_{S^1}
\phi_-^{i}=\eta^{i},\qquad \delta_{S^1} \eta^{i}=\hbar
\iota_{v_0}d\phi_-^i, \qquad \delta_{S^1}\theta^{i}=G_-^{i},\qquad
\delta_{S^1} G_-^{i}=\hbar \iota_{v_0}d\theta^{i}, \ee
$$
\delta_{S^1} \rho^i=-d\phi_+^i-\hbar \iota_{v_0}G_+^i, \qquad \delta_{S^1}
 \phi_+^i=\hbar \iota_{v_0}\rho^i,
 \qquad \delta_{S^1} G_+^i=d\,\rho^{i},
$$
where now the fields $\eta^i$ and $\theta^i$ are odd real zero-form
valued fields, $\rho^i$ are odd real one-form valued fields, $G_-^i$
are even real zero-form valued fields    and $G_+^i$ are even real two-form
valued fields. The action of the sigma model for the new real
structure is now given by
\be\label{linearTFTRone}
S=-\imath \sum_{j=1}^{\ell+1}\int_{D}\, \left((d\phi_+^j+\hbar
  \iota_{v_0}G_+^j)\wedge  *d\phi_-^{j}+\rho^j\wedge *d\eta^{j}
- \theta_jd\rho^j+ G_+^jG_-^{j}\right)
\ee
$$
+\sum_{i,j=1}^{\ell+1}
\int_{D}d^2z \sqrt{h}\left(-\frac{\pr^2 W_-(\phi_-)}{\pr
    \phi_-^i\pr\phi_-^j}
\eta^{i}\theta^{j}-\imath
\frac{\pr W_-(\phi_-)}{\pr\phi_-^i}\,G_-^{i}\right)
$$
$$
+\sum_{i,j=1}^{\ell+1}\int_{D}\left(-\frac{1}{2}
  \frac{\pr^2 W_+(\phi_+)}{\pr\phi_+^i\pr \phi_+^j}\rho^i\wedge \rho^j+
\frac{\pr W_+(\phi_+)}{\pr \phi_+^i}\,G_+^i\right)
-\frac{1}{\hbar}\int_{S^1=\pr D}\,d\sigma W_+(\phi_+).
$$
Here $W_+$ and $W_-$ are arbitrary independent regular functions on
$\IR^{\ell+1}$. Thus defined action is $\delta_{S^1}$-closed.

\begin{rem}
 Our choice of the real structure is such that the constructed
type B topological sigma model is a mirror dual
to the type A topological sigma model  considered in \cite{GLO1}.
In Section 3.3 we demonstrate that the mirror correspondence
applied to the type A topological  sigma models from \cite{GLO1}
leads to the real  structure of type  \eqref{newRS}.
 Note also that the construction of the
topological Yang-Mills theories using an  equivariant setting
\cite{W1} also leads
 to the non-standard real structure analogous to the one we use.
\end{rem}

In the following we consider the case of $W_-(\phi_-)=0$.  Thus we have
\be\label{linearTFTR}
S=-\imath \sum_{j=1}^{\ell+1}\int_{D}\, \left(
(d\phi_+^j+\hbar \iota_{v_0}G_+^j)\wedge *d\phi_-^{j}+\rho^j\wedge *d\eta^{j}
- \theta_jd\rho^j+ G_+^jG_-^{j}\right)
\ee
$$
+\sum_{i,j=1}^{\ell+1}\int_{D}\left(-\frac{1}{2}
  \frac{\pr^2 W_+(\phi_+)}{\pr\phi_+^i\pr \phi_+^j}\rho^i\wedge \rho^j+
\frac{\pr W_+(\phi_+)}{\pr \phi_+^i}\,G_+^i\right)
-\frac{1}{\hbar}\int_{S^1=\pr D}\,d\sigma W_+(\phi_+).
$$
Given an observable $\CO(z,\zb)$ on the disk $D$ we define its
correlation function as  a functional integral below
 \be
  \bigl\<\CO(z,\zb)\bigr\>_{W_+}\,:=\,
  \int\!D\mu\,\,\CO(z,\zb)\,e^{-S_*}\,,\\
  D\mu\,=\,\prod_{i=1}^{\ell+1}\,
  [D\phi_+^i][D\phi_-^i][D\eta^i][D\theta^i][D\rho^i][DG_+^i][DG_-^i]\,.
 \ee

\begin{lem} The following observable inserted at the center $z=0$  of the disk $D$
 \be\label{soakup}
\CO_*(0):=\CO_*(z,\zb)|_{z=0}\,=\,\prod_{i=1}^{\ell+1}\,
  \delta(\phi_-^i(z,\zb))\,\eta^i(z,\zb)|_{z=0}
 \ee
is $\delta_{S^1}$-invariant.
\end{lem}
\emph{Proof. }  We have
$$
  \delta_{S^1}\CO_*(z,\zb)\,=\,
  \sum_{m=1}^{\ell+1}\eta^m(z,\zb)
  \prod_{j\neq m}\delta(\phi_-^j)\prod_{i=1}^{\ell+1}\,\,\eta^i(z,\zb)
  $$
$$
  +\,\sum_{m=1}^{\ell+1}\prod_{j}\delta(\phi_{-}^j)(-1)^m
  \eta^1...\eta^{m-1}(\h\iota_{v_0}\eta^m)\eta^{m+1}....\eta^{\ell+1}\,.
$$
The first term is equal to zero  since for odd variables $\eta^2=0$.
The second term  vanishes since the
center of the disk $z=0$ is a fixed point of  the $S^1$-action  so that
$\iota_{v_0}(\eta^m)\bigr|_{z=0}=0$.
 $\,\Box$

\begin{te}
The correlation function of the observable
\eqref{soakup} in the type $B$ topological
$S^1$-equivariant linear sigma model \eqref{linearTFTR}
is given by
\be\label{finalex}
\<\CO_*(0)\>_{W_+}=\int_{\IR^{\ell+1}}\,\prod_{j=1}^{\ell+1}\,\,
dt^j\,\,\,e^{\frac{1}{\hbar}\,W_+(t)}.
\ee
\end{te}

\proof  Firstly we make an
integration over $G_-^i$:
$$
 \int[DG_-]\,\exp\Big\{\i\int_D\,
 \sum_{i=1}^{\ell+1}G_+^i(z)G_-^i(z)\Big\}\,=\,
 \prod_{i=1}^{\ell+1}\delta(G_+^i)\,.
$$
The integration over $G_+^j$ is then equivalent to the substitution
of $G_+^j=0$. Thus we should calculate the following functional integral:
 \be\label{Int2}
Z= \int[D\phi_+]\,[D\phi_-]\,\CO_1(0)\,\exp\Big\{\i\int_D\,
  \sum_{i=1}^{\ell+1}d\phi_+^i\wedge*d\phi_-^i-\frac{1}{\hbar}\int_{S^1}
d\sigma W_+(\phi_+)\Big\}\,Z_f(\phi_+)\,,
 \ee
where
 \be\nn
 Z_f(\phi_+)= \int\,[D\rho]\,[D\theta][D\eta]\,\,\CO_2(0)\,
  \exp\Big\{\i\int_D\,\sum_{i=1}^{\ell+1}\bigl(
  \rho^i\wedge*d\eta^i
  -\theta^i d\rho^i)+\frac{1}{2}\int_D\,\sum_{i,j=1}^{\ell+1}
  \frac{\pr^2W_+}{\pr\phi_+^i\pr\phi_+^j}\,\rho^i\wedge\rho^j\,\Big\}\,,
 \ee
and
$$
\CO_1(0)\,=\,\prod_{j=1}^{\ell+1}\delta(\phi_-^i(0)),\qquad
\CO_2(0)=\prod_{j=1}^{\ell+1}\eta^i(0).
$$
Let us first integrate  over $\theta$ in $Z_f$. We have
$$
Z_f(\phi_+)= \int\,[D\rho]\,[D\eta]\,\,\CO_2(0)\,\prod_{j=1}^{\ell+1}\delta(d\rho^j)
\exp\Big\{\i\int_D\,\sum_{j=1}^{\ell+1}\rho^j\wedge*d\eta^j+\frac{1}{2}\int_D\,
 \sum_{i,j=1}^{\ell+1} \frac{\pr^2W_+}{\pr\phi_+^i\pr\phi_+^j}\,\rho^i\wedge\rho^j
\,\Big\}\, .
$$
One-forms allow the following decomposition:
\be\label{oneformexp}
\rho^i=df^i_1+*df^i_2=\pr_z\bar{F^i}dz+\pr_{\zb} F^id\zb, \qquad F^i=f^i_1-\imath f^i_2.
\ee
It is easy to check  (using for example series expansions) that
for given $\rho^i$ the solutions $f_1$, $f_2$ of \eqref{oneformexp}
always exist and are unique up to  addition to $F^i$  a  holomorphic function.
Therefore we make the  following change of variables
$\rho^i \to (f^i_1,f^i_2)/\sim$ where the equivalence relation is
generated by  addition to $f_1^j$ and $f_2^j$
of real and imaginary parts of a holomorphic  function $g(z)$
 \be\label{eqrelone}
  f^i_1\,\sim\,f^i_1+{\rm Re}(g^i(z)),\qquad
  f^i_2\,\sim\,f^i_2+{\rm Im }(g^i(z)).
 \ee
Thus we have
$$
[D\rho]=\frac{[Df_1]\,[Df_2]}{[Dg]}\,Jac_1^{-1}\,,
$$
where Jacobian is given by the determinant of the operator
$$
(d\oplus *d):\,\,(f^i_1,f^i_2)\rightarrow \rho^i=df^i_1+*d f^i_2,
$$
acting $\CA^0_{orth}\subset\CA^0(D)$ orthogonal
to its kernel. We define a determinant of an  operator acting
between different spaces as a square root of the determinant of the
product of the operator and its conjugated
$$
Jac_1=|\det{}'_{\CA_{orth}^0\oplus \CA_{orth}^0} (d+*d)|:=
\left(\det{}'_{\CA_{orth}^0\oplus \CA_{orth}^0}
  (d+*d)^2\right)^{\frac{1}{2}}=\det{}'_{\CA_{orth}^0}\Delta_0,
$$
where $\Delta_0=(d+d^*)^2$ acting in the space of  functions $\CA_0$. We have
 $$\delta(d\rho^i)=\delta(d(df^i_1+*df^i_2))=\delta(d*df^i_2),$$ and
 thus
$$
Z_f(\phi_+)= \int\,[D\eta]\,\frac{[Df_1]\,[Df_2]}{[Dg]}\,
\,\frac{1}{\det{}'_{\CA_{orth}^0} \Delta_0}\,
\CO_2(0)\,\prod_{i=1}^{\ell+1}\delta(d*d\,\,f^i_2)
$$
$$
 \times\,\exp\Big\{
 \i\int_D\,\sum_{i=1}^{\ell+1}(df^i_1+*d\,\,f^i_2)\wedge*d\eta^i
 \,+\frac{1}{2}\int_D\,
 \sum_{i,j=1}^{\ell+1} \frac{\pr^2W_+}{\pr\phi_+^i\pr\phi_+^j}\,
(df^i_1+*d\,\,f^i_2)\wedge (df^j_1+*d\,\,f^j_2)\Big\}\, .
$$
Let us fix a representative for the equivalence relation
\eqref{eqrelone} by the condition that $f^i_2$ is in the subspace
orthogonal to the space of harmonic functions on the disk.  This
leaves a freedom to add to $f^i_1$  a real constant (indeed ${\rm
Im}(g^i(z))=0$ implies $g^i(z)=a^i\in\IR$).
We denote  by $[Df_1]'$ the induced measure on this subspace. The
integration over $f^i_2$ gives
$$
Z_f(\phi_+)= \int\,[D\eta]\,[Df_1]'\,
\,\,\, \CO_2(0)\,\exp\Big\{\i\int_D\,\sum_{i=1}^{\ell+1}df^i_1\wedge*d\eta^i+
\,\frac{1}{2}\int_D\,
 \sum_{i,j=1}^{\ell+1} \frac{\pr^2W_+}{\pr\phi_+^i\pr\phi_+^j}\,
df^i_1\wedge df^j_1\Big\}\,,
$$
where the determinant in the denominator is canceled by the determinant
appearing from the integration of the delta-function.

We split the space of functions $\CA^0(D)$ on a disk on 
the space $\CA^0_h$ of harmonic
functions and the space $\CA^0_N$ of  functions that have zero normal derivative on the
boundary:
 \be\nn
  f^i=f^i_{h}+f^i_N\,,
  \qquad
  f^i_h\in\CA^0_h,\,\quad
  f^i_N\in\CA^N\,,\\
  \Delta_0f^i_h=0\,,
  \hspace{1.5cm}
  \pr_nf^i_N|_{S^1}=0\,.
 \ee
The subspace $\CA^0_h$ can be identified with the space ${\rm
Fun}(S^1)$ of functions on the boundary $S^1=\pr D$. This is not an
orthogonal decomposition with respect to the natural scalar product
on the space of functions on the disk. Thus we have a non-trivial
Jacobian in the integration measure:
$$
 [Df]=[Df_h]\,[Df_N]\,Jac_2^{-1},
$$
which is a some constant. Note that the following relation holds:
$$
\int_D \,\sum_{i=1}^{\ell+1}\,df^i_1\wedge * d \eta^i
=\int_D\,\sum_{i=1}^{\ell+1}\, \eta^i_N\,*\Delta f^i_{1,N} -
\int_{S^1}\,\sum_{i=1}^{\ell+1}\,\eta^i_h*df^i_{1,h}.
$$
Taking  integral over  $\eta^i_{1,N}$ and
$\eta^i_{1,h}$   we obtain
$$
Z_f(\phi_+)= \frac{1}{Jac_2}\,\int\,[D f_1]'\,\,\,
\prod_{i=1}^{\ell+1}\delta(\Delta_0f^i_{1,N})\,\delta(*df^i_{1,h})
$$
$$
 \times \,\exp\Big\{\frac{1}{2}\int_D\,
 \sum_{i,j=1}^{\ell+1} \frac{\pr^2W_+}{\pr\phi_+^i\pr\phi_+^j}\,
 d(f^i_{1,N}+f_{1,h})\wedge d(f^j_{1,N}+f_{1,h})\Big\}\,
 =\,\frac{1}{Jac_2^2}\det{}'_{\CA^0_N}\,\Delta_0\,\,
 \det{}'_{{\rm Fun}(S^1)}(*d)\,.
$$
Now let us  calculate the functional integral \eqref{Int2}. The
calculation is basically the same as in the case of $Z_f$. The only
difference (apart of the fact   that Jacobins and determinants
appear inverse) is that  the integral over  constant mode of
$\phi_{-}^j$ is present and is eaten up by the delta-function
insertion. On the other hand the integral over  constant mode of
$\phi_+^j$ remains. Taking into account the cancelation of the
Jacobians and determinants for fermions and bosons the total
integral is equal to
$$
Z=\int_{\IR^{\ell+1}}\,\prod_{j=1}^{\ell+1}\,\,
dt^j\,\,\,e^{\frac{1}{\hbar}\,W_+(t)},
$$
where $t^j$ are constant  modes of the fields $\phi^j_+$. $\Box$

\begin{cor}\label{Bcor}
The correlation function of the observable
\eqref{soakup} in the type $B$ topological
$S^1$-equivariant linear sigma model \eqref{linearTFTR}
   with the  superpotential
\be\label{superpot}
W^{(0)}_+(\phi_+)=\sum_{j=1}^{\ell+1}(\lambda_j\phi^j_+-e^{\phi_+^j}),
\ee
is given by the  following  product of  the $\Gamma$-functions
\be
\<\CO_*(0)\>_{W^{(0)}_+}=\prod_{j=1}^{\ell+1}\,
\hbar^{\frac{\lambda_j}{\hbar}}\,\Gamma\left(\frac{\lambda_j}{\hbar}\right).
\ee
\end{cor}
\proof Using the result of the previous Theorem for the superpotential
\eqref{superpot} we straightforwardly have
$$
\<\CO_*(0)\>_{W^{(0)}_+}=\int_{\IR^{\ell+1}}\,\prod_{j=1}^{\ell+1}
dt^j\,\,\,\,e^{\frac{1}{\hbar}\sum_{j=1}^{\ell+1}
(\lambda_jt^j-e^{t^j})}=
\prod_{j=1}^{\ell+1}\,\hbar^{\frac{\lambda_j}{\hbar}}\,
\Gamma\left(\frac{\lambda_j}{\hbar}\right).
$$
$\Box$

The expression (2.15) is equivalent to the one obtained in type $A$ topological
sigma model considered in \cite{GLO1}. The coincidence of a
particular correlation functions in type $A$ model considered in
\cite{GLO1} and the correlation function from Corollary \ref{Bcor}
is a manifestation of the mirror symmetry between two underlying
sigma models. Without taking into account  the involved
$S^1$-equivariance, the mirror correspondence between the 
two models follows  from the results of \cite{HV}. In particular the
exponential terms in the superpotential \eqref{superpot} are 
attributed to the summation over instantons in type $A$ sigma model.
In the following Section we provide heuristic arguments for the
mirror symmetry between the topological theory considered in this
note  and  the one considered in \cite{GLO1}.

\section{On equivalence of $A$ and $B$ topological sigma models}

As it was demonstrated in the previous Section the Euler integral
representation of the $\Gamma$-function
\be\label{Euler}
\Gamma(s)=\int_{-\infty}^{+\infty}\,dx\,e^{xs}\,e^{-e^{x}},\qquad {\rm
  Re}(s)>1,
\ee naturally arises as a particular correlation function in  a
certain $S^1$-equivariant type $B$  topological sigma model on
the disk $D$. In \cite{GLO1} it was argued that this integral
representation is dual to the representation of  the
$\Gamma$-function as an equivariant symplectic volume of an
infinite-dimensional space. The natural framework for this duality
is a mirror symmetry. Below  we establish a direct relation of the
Euler integral representation \eqref{Euler} of the $\Gamma$-function
with the representation of the $\Gamma$-function as an equivariant
symplectic volume of an infinite-dimensional space proposed in
\cite{GLO1}. We also  discuss an explicit mirror map between 
the type $A$ equivariant topological linear sigma model considered
in \cite{GLO1} and the type $B$ equivariant topological sigma model
considered in the previous Sections. Finally we elucidate the
appearance of the non-standard  real structure \eqref{newRS} in
a simple example of the mirror map for a sigma model  on $\IP^1$
with  the target space being an infinite cylinder $\IC^*=\IR\times
S^1$.

\subsection{Fixed point calculation of equivariant volume}

In this Subsection we derive the Euler integral representation
of the Gamma-function \eqref{Euler} applying the Duistermaat-Heckman
fixed point formula to the infinite-dimensional integral representation
for the Gamma function proposed in \cite{GLO1}.

Let us start with recalling  the functional integral representation
of the $\Gamma$-function as an equivariant symplectic volume from
\cite{GLO1}. Let $\CM(D,\IC)$ be a  space of holomorphic maps of the
disk $D=\{z\in \IC|\,\,|z|\leq 1\}$ into the complex plane 
$\IC$. An element of $\CM(D,\IC)$ can be described  as  a complex
function $\varphi(z,\zb)$ on $D$,
 satisfying the equation
\be \pr_{\zb}\varphi(z,\zb)=0. \ee We denote the complex conjugated
function by $\bar{\varphi}(z,\zb)$. Define a symplectic form on the
space $\CM(D,\IC)$  as follows
 \be\label{Bsympf}
  \Omega=\frac{\imath}{4\pi} \,\int_0^{2\pi}\,\delta
  \varphi(\sigma)\wedge \delta\bar{\varphi}(\sigma)\,d\sigma,
 \ee
where $\varphi(\sigma)$, $\bar{\varphi}(\sigma)$ are restrictions of
$\varphi(z,\zb)$, $\bar{\varphi}(z,\zb)$ to  the boundary $\pr
D=S^1$ and $\sigma$ is a coordinate on the boundary such that
$\sigma\sim \sigma+2\pi$.  The symplectic form \eqref{Bsympf} is
invariant with respect to the action of the group $S^1$ of loop
rotations and to the action of $U(1)$  induced from the standard
action of $U(1)$ on $\IC$ \be\label{U1act} \varphi(z)\longrightarrow
e^{\imath \alpha}\varphi(z), \qquad
\bar{\varphi}(\zb)\longrightarrow e^{-\imath
  \alpha}\bar{\varphi}(\zb), \qquad e^{\imath
  \alpha}\in U(1),
\ee
\be
\varphi(z)\longrightarrow \varphi(e^{\imath \beta}z),
\qquad \bar{\varphi}(\zb)\longrightarrow \bar{\varphi}(e^{-\imath
  \beta}\zb), \qquad e^{\imath
  \beta}\in S^1.
\ee Let $\hbar$ and $\lambda$ be  generators of the Lie algebras of
$S^1$ and $U(1)$ correspondingly.   The action of $S^1\times U(1)$
on $(\CM(D,\IC),\Omega)$ is Hamiltonian and  the corresponding
momenta are given by
 \be\label{Momenta}
  H_{S^1}=-\frac{\imath}{4\pi}\int_0^{2\pi}\, \bar{\varphi}
  (\sigma)\pr_\sigma \varphi(\sigma)\, \,d\sigma,\qquad
  H_{U(1)}=\frac{1}{4\pi} \int_0^{2\pi}\,
  |\varphi(\sigma)|^2\,\,d\sigma.
 \ee
The $S^1\times U(1)$-equivariant volume  of $\CM(D,\IC)$ is defined
formally as follows \cite{GLO1}. Let $\chi(z,\zb)$ and
$\bar{\chi}(z,\zb)$ be a pair of complex conjugated odd functions
 satisfying the equations
\be
\pr_{\zb}\chi(z,\zb)=0,\qquad
\pr_{z}\bar{\chi}(z,\zb)=0.
\ee
The functions  $(\chi(z,\zb), \bar{\chi}(z,\zb))$
can be considered as a section of the odd tangent bundle
$\Pi T\CM(D,\IC)$ to $\CM(D,\IC)$. Using the standard correspondence
between differential forms on a manifold $X$ and the functions on the odd
tangent bundle $\Pi TX$ one can write down the symplectic form \eqref{Bsympf}
as follows:
$$
\Omega=\frac{\imath}{4\pi}\int_0^{2\pi}d\sigma\,\chi(\sigma)\,\bar{\chi}(\sigma).
$$
Below  we  freely use the equivalence between differential forms and
functions on superspaces without further notice.

The  $S^1\times U(1)$-equivariant volume of the space of holomorphic
maps $\CM(D,\IC)$ is given by the following functional integral:
\be\label{intSI} Z(\lambda,\hbar,\mu)= \int_{\Pi
T\CM(D,\IC)}\,dm(\varphi,\chi)\,\,\, \,e^{\mu(\lambda
H_{U(1)}+\hbar\,H_{S^1}+\Omega)},\qquad {\rm Re}(\mu)< 0, \ee where
$H_{S^1}$, $H_{U(1)}$ are given by \eqref{Momenta}, and
$dm(\varphi,\chi)$ is a canonical integration measure on the
superspace $\Pi T\CM(D,\IC)$ defined in \cite{GLO1}. The integral
\eqref{intSI} is an  infinite-dimensional Gaussian integral and is  understood
 using the zeta-function regularization. Note that  in general,  regularized
infinite-dimensional integrals depend on auxiliary  parameters
defined by a particular choice of a regularization scheme.
For the integral \eqref{intSI} this leads to the following 
general dependence on a regularization scheme \cite{GLO1}:
\be\label{resultA}
Z(\lambda,\hbar,\mu)=A(\mu)\,B(\mu)^{\frac{\lambda}{\hbar}}\,\,
\Gamma\left(\frac{\lambda}{\hbar}\right), \ee 
where $A(\mu)$ and $B(\mu)$ are some $\lambda$-independent 
functions. Thus taking into account the dependence  on a
choice of a regularization scheme it is natural to consider the
$S^1\times U(1)$-equivariant volume of the space of holomorphic maps
$\CM(D,\IC)$ (and thus in particular the  Gamma-function) as a $\IR^*\times
\IR_+$-torsor. The regularization scheme we use below leads to a
particular choice of $A$ and $B$. 

In \cite{GLO1}  the integral \eqref{intSI} was expressed in terms
of infinite-dimensional determinant and
no obvious relation with the Euler integral representation
\eqref{Euler} was given. Below we consider a heuristic  derivation of
\eqref{resultA} using an infinite-dimensional version of the
Duistermaat-Heckman fixed point formula \cite{DH}. In this derivation
the Euler integral representation  \eqref{Euler} appears in a natural way.

To proceed  let us first recall a construction
of a projective space $\IP^N$  as the
 Hamiltonian reduction of a symplectic manifold $(\IC^{N+1},\omega_{\IC^{N+1}})$
by the  Hamiltonian action of the group $U(1)$. Here the
symplectic form $\omega_{\IC^{N+1}}$ is given by \be
 \omega_{\IC^{N+1}}\,
 =\,\frac{\imath}{2}\sum_{j=1}^{N+1}\,\,dz_j\wedge d\zb_j,
 \ee
and  the $U(1)$ action
 \be\label{Uoneact}
  e^{\imath \alpha}:\,\,\,z_j\longrightarrow e^{\imath
  \alpha}\,z_j,\qquad e^{\imath \alpha}\in U(1), \qquad j=1,\ldots ,N+1,
 \ee
is generated by the vector field
$$
 v\, =\,\sum_{i=1}^{N+1}\imath\Big\{z_i\frac{\pr}{\pr z_i}\,
 -\,\zb_i\frac{\pr}{\pr\zb_i}\Big\}.
 $$
The momentum $H_{U(1)}$ corresponding to the Hamiltonian action
\eqref{Uoneact} is defined by the equation
$\iota_v\omega=-dH_{U(1)}$ and is given by  $
H_{U(1)}=\frac{1}{2}\sum_{j=1}^{N+1}\,|z_j|^2$. Projective space
$\IP^N$  can be realized as  a Hamiltonian reduction of
$(\IC^{N+1},\omega_{\IC^{N+1}})$ by  $U(1)$
 \be\label{hmom}
  \IP^N=\Big\{z\in\IC^{N+1}\,\Big|\,H_{U(1)}(z,\zb)=\frac{1}{2}r^2\Big\}\Big/U(1)\,,
  \qquad r\in \IR.
 \ee
Thus constructed $\IP^N$ has  a canonical symplectic structure
$\omega_{\IP^N}$ proportional to the Fubini-Study form.  In terms of
inhomogeneous coordinates $w_j=z_j/z_{N+1}$, $z_{N+1}\neq 0$ it is
given by
 \be\label{FS}
  \omega_{\IP^N}=\frac{\imath r^2}{2}
  \frac{(1+\sum_{i=1}^N|w_i|^2)\sum_{j=1}^Ndw_j\wedge d\wb_j
  -\sum_{i,j}^Nw_i\wb_jdw_j\wedge d\wb_i}{(1+\sum_{i=1}^N|w_i|^2)^2}.
 \ee
The symplectic space $(\IC^{N+1},\omega_{\IC^{N+1}})$ allows also the
Hamiltonian action of the group $U(1)^{N+1}$
 \be
  z_i\,\longmapsto\,z_ie^{\i\alpha_i}\,,
  \hspace{1.5cm}
  e^{\i\alpha_i}\in U(1)_i\,,
  \hspace{1cm}
  i=1,\ldots,N+1\,,
 \ee
generated by  vector fields
$$
 v_i\,=\,\imath\Big\{z_i\frac{\pr}{\pr z_i}\,
 -\,\zb_i\frac{\pr}{\pr\zb_i}\Big\}\,,
 \hspace{2.5cm}
 i=1,\ldots,N+1\,.
$$
Solving the equations $\iota_{v_i}\omega_{\IC^{\ell+1}}=-dH_i$ we find the
 corresponding momenta
$$
 H_i\,=\,\frac{1}{2}|z_i|^2\,,
 \hspace{2.5cm}
 i=1,\ldots,N+1\,.
$$
The action of $U(1)^{N+1}$  descents to the Hamiltonian action on  $(\IP^N,
\omega^{\IP^N})$ with the corresponding momenta
 \be\label{PH1}
  H^{\IP^N}_j=\frac{r^2}{2}\,\frac{|w_j|^2}{1+\sum_{j=1}^N|w_j|^2}\,,
  \qquad j=1,\ldots N,
 \ee
and
 \be\label{PH2}
  H^{\IP^N}_{N+1}=\frac{r^2}{2}\,\frac{1}{1+\sum_{j=1}^N|w_j|^2}.
 \ee

\begin{lem}
 The following identity holds:
 \be\label{PellVolumeForms}
\frac{1}{2\pi \mu}  \int_{\IC^{N+1}}\,\,
  \delta\Big(H_{U(1)}\,-\,r^2/2\Big)\,e^{\mu(\omega_{\IC^{N+1}}+\sum_{j=1}^{N+1}
\lambda_j H_j)}
  =\,\int_{\IP^N}\,\,e^{\mu(\omega_{\IP^N}+\sum_{j=1}^{N+1}\lambda_j
H^{\IP^N}_j)}\,,
 \ee
where $\omega_{\IP^N}$ is given by \eqref{FS} and
the reduced Hamiltonians $H^{\IP^N}_j$ are given by \eqref{PH1} and \eqref{PH2}.
\end{lem}
\emph{Proof.}  Let us introduce new variables
$w_j=z_j/z_{N+1},\,j=1,\ldots,N$ and $t=|z_{N+1}|^2$,
$\theta=\frac{1}{2\imath}\ln\frac{z_{N+1}}{\zb_{N+1}}$, so that
$z_{N+1}=\sqrt{t}\,e^{\imath\theta}$. Then we have
 \be
\frac{\mu^{N}}{2\pi}\left(\frac{\imath }{2}\right)^{N+1}\,
  \int_{\IC^{N+1}}\bigwedge_{i=1}^{N+1}
  dz_i\wedge d\zb_i\,\delta\Big(\frac{1}{2}
\sum_{i=1}^{N+1}|z_i|^2-\frac{r^2}{2}\Big)e^{\mu\sum_{j=1}^{N+1}\lambda_j H_j}\\\nn
  =\,\frac{\mu^N\,r^{2N}}{2\pi}\left(\frac{\imath }{2}\right)^{N}
\int_0^{2\pi} d\theta\int_0^{\infty}dt\,t^{N}\int_{\IC^{N}}
  \frac{\bigwedge\limits_{n=1}^{N}(dw_n\wedge d\wb_n)}{
  1+\sum|w_n|^2}\,
  \delta\Big(t-\frac{r^2}{1+\sum|w_n|^2}\,\Big)e^{\mu\sum_{j=1}^{N+1}\lambda_j H_j}\\
  =\mu^N r^{2N}\left(\frac{\imath }{2}\right)^{N}
\int_{\IC^{N}}
  \frac{\bigwedge\limits_{n=1}^{N}(dw_n\wedge d\wb_n)}
  {\bigl(1+\sum|w_n|^2\bigr)^{N+1}}\,e^{\mu\sum_{j=1}^{N+1}\lambda_j
H^{\IP^N}_j}.
 \ee
Taking into account that
$$
\frac{\omega_{\IP^N}^N}{N!}= r^{2N}\left(\frac{\imath }{2}\right)^{N}
\frac{\bigwedge\limits_{n=1}^{N}(dw_n\wedge d\wb_n)}
  {\bigl(1+\sum|w_n|^2\bigr)^{N+1}},
$$
we obtain the identity \eqref{PellVolumeForms}. $\Box$

We  shall use an infinite-dimensional analog of the identity
\eqref{PellVolumeForms} to calculate the integral \eqref{intSI}.
Let us  rewrite the integral \eqref{intSI}  as follows:
\be\label{infolone}
Z(\lambda,\hbar,\mu)= \int_{-\infty}^{+\infty} dt\,e^{\mu\lambda
  t}\,Z_t(\hbar,\mu),\qquad
Z_t(\hbar,\mu)=\int_{\CM(D,\IC)} e^{\mu(\hbar
H_{S^1}+\Omega)}\,\,\delta(t-H_{U(1)}). \ee Now taking into account
\eqref{PellVolumeForms} we can interpret $Z_t(\hbar,\mu)$ as an
integral over the infinite-dimensional projective space
$\IP\CM(D,\IC)$ \be\label{infvol} Z_t(\hbar,\mu)=2\pi
\mu\,\int_{\IP\CM(D,\IC)}e^{\mu(\hbar   \tilde{H}_{S^1}+\Omega(t))},
\ee where $\Omega(t)$ is an induced symplectic form on
$\IP\CM(D,\IC)$ and $\tilde{H}_{S^1}$ is a  momentum corresponding
to the $S^1$-action on $\IP\CM(D,\IC)$. We should stress that
the integral in  \eqref{infvol} is an infinite-dimensional one  and
thus requires a proper regularization which will be discussed below.

To calculate the integral \eqref{infvol} we use
an infinite-dimensional version of the  Duistermaat-Heckman formula
\cite{DH} (for a detailed introduction into the subject see e.g. \cite{Au}).
 Let $M$ be a $2N$-dimensional  symplectic manifold with
the  Hamiltonian action of  $S^1$ having only isolated fixed points.
Let $H$ be the corresponding momentum.
The tangent space $T_{p_k}M$ to a fixed point $p_k\in M^{S^1}$ has a
natural action of $S^1$. Let $v$  be  a generator
of ${\rm Lie} (S^1)$ and let $\hat{v}$ be  its action on
$T_{p_k}M$. Then the following identity holds:
 \be\label{DH}
  \int_M\,e^{\mu(\hbar H+\omega)}=\sum_{p_k\in M^{S^1}}\,
  \frac{e^{\mu\hbar H(p_k)}}{\det_{T_{p_k}M}\hbar\hat{v}/2\pi}\,\,.
 \ee
Let us formally apply \eqref{DH} to the integral \eqref{infvol}.
A set of fixed points of $S^1$ acting on $\IP\CM(D,\IC)$   can be  easily
found using linear coordinates on $\CM(D,\IC)$ (considered as
homogeneous coordinates on $\IP\CM(D,\IC)$). Let $\varphi(z)$ be a
holomorphic map of $D$ to $\IC$. It represents an $S^1$-fixed point
on $\IP\CM(D,\IC)$ if rotations by $S^1$ can be
compensated  by an action of  $U(1)$
\be\label{fpcond}
e^{\imath \alpha(\beta)}\varphi(e^{\imath \beta}z)=\varphi(z), \qquad
\beta\in [0,2\pi].
\ee
It is easy to see that solutions of \eqref{fpcond}
are enumerated by non-negative integers and are given by
\be\label{fixedpt}
 \varphi^{(n)}(z)=\varphi_n z^n,\qquad \varphi_n\in \IC^*\, \quad n\in \IZ_{\geq 0}.
\ee
The tangent space to $\CM(D,\IC)$ at an  $S^1$-fixed point $\varphi^{(n)}$
has natural linear coordinates $\varphi_m/\varphi_n$, $m\in \IZ_{\geq 0},
m\neq  n$ where coordinates $\varphi_k$, $k\in \IZ_{\geq 0}$  are defined by the series
expansion  of $\varphi\in \CM(D,\IC)$
$$
\varphi(z)=\sum_{k=0}^\infty\varphi_kz^k.
$$
After identification of $\hbar$ in \eqref{infvol} with a
generator of ${\rm Lie}(S^1)$ its action on  the tangent space
at the fixed point
is given by a  multiplication of each  $\varphi_m/\varphi_n$  on $(m-n)$.
Thus to define an analog of the  denominator in the right hand side of
the  Duistermaat-Heckman formula \eqref{DH} one should provide a
meaning to the infinite product $\prod_{m=0,m\neq n}^{\infty}\hbar(m-n)/2\pi$.
We use a $\zeta$-function regularization  (see e.g. \cite{H} and also  Appendix in
\cite{GLO1})
 \be\label{regprod}
  \ln \Big[\prod_{{m\in\IZ_{\geq 0},},m\neq n}\,
  \frac{\hbar}{2\pi}(m-n)\Big]_a\,:=\,-\frac{\pr}{\pr s}\left.\left(
  \sum_{m=1}^n\frac{e^{-\imath \pi s}}{(a\hbar m/2\pi )^s}
  +\sum_{m=1}^{\infty}\frac{1}{(a \hbar m/2\pi )^s}\right)\right|_{s\to0},
\ee
where $a$ is a normalization multiplier.
The introduction of $a$ is to take into account a multiplicative
anomaly $\det (A B)\neq \det A\cdot \det B$ appearing for generic
operators $A$ and $B$. We specify $a$ at the final step of the calculation of
\eqref{infvol}.

\begin{lem} The regularized product \eqref{regprod} is given by
 \be\label{prodd}
  \frac{1}{\left[\prod_{m\in\IZ_{\geq 0}, m\neq n}\hbar (m-n)/2\pi \right]_a}
  =(-1)^n\frac{(a \hbar/2\pi )^{-n}}{n!}\frac{\sqrt{a\hbar}}{2\pi}.
 \ee
\end{lem}

\proof   Using the Riemann $\zeta$-function
$$
\zeta(s)=\sum_{n=1}^{\infty}\frac{1}{n^s},
$$
one can express the right hand side of \eqref{regprod}  as follows:
$$
 \ln \Big[\prod_{{m\in\IZ_{\geq 0},},m\neq n}\,
 \frac{\hbar}{2\pi}(m-n)\Big]_a=
 (\zeta(0)+n)\ln a \hbar/2\pi  +\ln n!-\zeta'(0)+\imath \pi n.
$$
Taking into account  $\zeta(0)=-\frac{1}{2}$ and
$\zeta(0)'=-\frac{1}{2}\ln 2\pi$ we obtain \eqref{prodd}. $\Box$

Let us now calculate the difference of the values of  $S^1$-momentum map
$\tilde{H}_{S^1}$ at two $S^1$-fixed points $\varphi^{(n)}, \varphi^{(0)}\in
  \IP\CM(D,\IC)$. Consider an embedded  projective line $\IP^1\subset\IP\CM(D,\IC)$,
containing $\varphi^{(n)}$ and $\varphi^{(0)}$. Let us choose homogeneous
coordinates $[z_0:z_1]$ on $\IP^1$ such that $\varphi^{(0)}=[1:0]$ and
$\vp^{(n)}=[0:1]$. The action of $S^1$ on $\IP\CM(D,\IC)$
descends to the embedded  $\IP^1$ via the  vector field
\be\label{descentvec}
 V\,=\,\i n\,\Big\{w\frac{\pr}{\pr
   w}\,-\,\wb\frac{\pr}{\pr\wb}\Big\}\,,\qquad w=z_1/z_0.
\ee
The pull back of the symplectic form $\Omega(t)$ is
given by
$$
 \omega_{\IP^1}\,=\,\imath t\frac{dw\wedge d\wb}{(1+|w|^2)^2}\,.
$$
The action of the vector field \eqref{descentvec} on $\IP^1$  is the
Hamiltonian one. Let
$H^{(n)}_{S^1}$ be the corresponding momentum   given by a
restriction of the momentum $\tilde{H}_{S^1}$ for $S^1$-action
$\IP\CM(D,\IC)$. From the definition of the momentum map  we have
 \be
  H^{(n)}_{S^1}(\vp^{(n)})\,-\,H^{(n)}_{S^1}(\vp^{(0)})\,
  =\,\int_{[1:0]}^{[0:1]}dH^{(n)}_{S^1}\,
  =\,-\int_{[1:0]}^{[0:1]}\iota_{V}\omega_{\IP^1}\,.
 \ee
A momentum defined as a solution of the equation $i_V\omega=-dH$ is
unique up an   additive constant. To fix this constant
we normalize the momentum $\tilde{H}_{S^1}(\varphi)$ so that
$H_{S^1}(\varphi^{(0)})=0$. Thus  we obtain the following:
 \be\label{CriticalValues}
  H^{(n)}_{S^1}(\vp^{(n)})\, =\,nt\int_{[1:0]}^{[0:1]}
  \frac{wd\wb+\wb dw}{(1+|w|^2)^2}\, =\,-
nt\Big[\frac{1}{(1+|w|^2)}\Big]_0^{\infty}\,
  =\,nt\,.
\ee
Substituting \eqref{CriticalValues} and \eqref{prodd}
into \eqref{DH} for $M=\IP\CM(D,\IC)$ we obtain
 \be
  Z_t(\h,\mu)\,=2\pi \mu\,\sqrt{\frac{a \hbar}{(2\pi)^2}}\,
  \,\sum_{n=0}^{\infty}\,(-1)^n\,\frac{e^{nt\mu\h}}{(a \hbar/2\pi )^n\,n!}\,
  =\,\mu\sqrt{a\hbar}\,
  \exp\Big\{-\frac{2\pi }{a \hbar}\,e^{\mu\h t}\Big\}\,,
 \ee
where the dependence on the normalization constant $a$ reflects an ambiguity of the
regularized infinite-dimensional integral. Taking into account
\eqref{infolone}, the regularized $S^1\times U(1)$-equivariant
symplectic volume of $\CM(D,\IC)$ is given by
 \be\label{Intone}
  Z_{\rm reg}(\la,\,\h,\,\mu)\,=\,\int_0^{\infty}dt\,
  e^{\mu \la t}Z_t(\h,\, \mu)\,
  =\,\mu\sqrt{a \hbar}\,\int_0^{\infty}dt\,e^{\mu\la t}\,
  \,e^{-\frac{2\pi}{a \hbar}e^{\mu\h t}}\\
=\left(\frac{a}{\hbar}\right)^{1/2}\,
\left(\frac{a\hbar}{2\pi}\right)^{\frac{\lambda}{\hbar}}
\int^{+\infty}_{-\ln(a\hbar/2\pi)}du\,e^{\frac{\lambda}{\hbar}u}\,e^{-e^u},
 \ee
where $u=\mu\h t-\ln(a\hbar/2\pi)$. To get rid of the renormalization ambiguity
we take the limit $a\to +\infty$ in the following way:
 \be\label{Inttwo}
  Z(\la,\h)\,
  =\,\lim_{a\to +\infty} \left(\frac{a}{\hbar}\right)^{-1/2}\,
\left(\frac{a}{2\pi}\right)^{-\frac{\lambda}{\hbar}}
 Z_{\rm reg}(\CM;\,\la,\,\h)\,
  =\hbar^{\frac{\lambda}{\hbar}}\,\Gamma\Big(\frac{\la}{\h}\Big)\,.
 \ee
Thus we show that the formal application of the
Duistermaat-Heckman formula to the infinite-dimensional integral
\eqref{intSI} in the form \eqref{infolone}
leads to the Euler integral representation \eqref{Euler} of the
$\Gamma$-function and reproduces the results of Section 2.

\subsection{On explicit mirror  map for the target space $\IC$}

In this Subsection we consider  an explicit mirror map of the type $A$
topological sigma model considered in \cite{GLO1} to  the type $B$
topological sigma model considered in Section 1.

In the previous Sections  we take into account the action
\eqref{U1act} of $U(1)$ on the symplectic space
 $(\CM(D,\IC),\Omega)$  of holomorphic
maps of the disk $D$ into the complex plane $\IC$.
Now we introduce a larger infinite-dimensional group acting on $(\CM(D,\IC),\Omega)$
in a Hamiltonian way.
The space $(\CM(D,\IC),\Omega)$ supports the  Hamiltonian action of a
commutative Lie algebra $\CG={\rm Map}(S^1,\IR)$
of real functions on $S^1$  given by
$$
\alpha\cdot \varphi(\sigma)=
\imath \left[\alpha(\sigma) \varphi(\sigma)\right]_+,\qquad
\alpha\cdot  \bar{\varphi}(\sigma)
=-\imath \left[\alpha(\sigma) \bar{\varphi}(\sigma)\right]_-,
$$
where $\alpha(\sigma)\in \CG$ and
$\varphi(\sigma)$, $\bar{\varphi}(\sigma)$ are restrictions of
 $\varphi(z)$, $\bar{\varphi}(\zb)$ to the boundary $S^1=\pr D$. The
 projectors $[\,\,\,]_{\pm}$  are defined as follows:
$$
[e^{\imath n\sigma}]_+=e^{\imath n\sigma},\quad n\geq 0,\qquad
[e^{\imath n\sigma}]_+=0,\quad n<0,\qquad [e^{\imath
  n\sigma}]_-=e^{\imath n\sigma}-[e^{\imath n\sigma}]_+\,.
$$
Given a Hamiltonian action of $\CG$ one can define corresponding momentum map
of $\CM(D,\IC)$ into the dual to the Lie algebra $\CG$.
The value of the momentum on the element $\alpha(\sigma)$ of
the Lie algebra $\CG$ is given by
\be\label{momentumOne}
H_{\CG}(\alpha)=\int_0^{2\pi}d\sigma\,\,
\alpha(\sigma)\,H_{\CG}(\bar{\varphi}(\sigma),\varphi(\sigma)),\qquad
H_{\CG}(\bar{\varphi}(\sigma),\varphi(\sigma))=\frac{1}{4\pi}|\varphi(\sigma)|^2.
\ee
Note that the subalgebra  $\mathfrak{u}(1)\subset \CG$ corresponding
to  $\alpha(\sigma)=const$ coincides with the  Lie algebra of
the group $U(1)$ considered in the previous Subsection. The momenta \eqref{momentumOne}
motivate an introduction of a new parametrization of $\CM(D,\IC)$
$$
\varphi(\sigma)=\tau^{1/2}(\sigma)\,e^{\imath \phi(\sigma)},
\qquad
\bar{\varphi}(\sigma)=\tau^{1/2}(\sigma)\,e^{-\imath \phi(\sigma)},
$$
and thus
\be\label{newpar}
\tau(\sigma)=|\varphi(\sigma)|^2,\qquad \phi(\sigma)=-\frac{\imath}{2}
\ln\left(\frac{\varphi(\sigma)}{\bar{\varphi}(\sigma)}\right).
\ee
Note that  thus defined $\tau(\sigma)$ is constraint by  the
condition to be a restriction to the boundary $S^1$
of the square module of a holomorphic function on $D$. Also let us stress
that $\phi(\sigma)$ given by \eqref{newpar} is not single-valued. Indeed let
$\varphi^{(n)}(z)=p_n(z)\varphi^{(0)}(z)$ be a holomorphic function on $D$ such
that $p_n(z)=\prod_{j=1}^n(z-a_j)$, $a_j\in D$  is a polynomial of degree $n$
and $\varphi^{(0)}(z)$ is a holomorphic function without zeroes inside
$D$. Then we have for the corresponding function $\phi(\sigma)$
\be\label{monodromy}
\phi^{(n)}(\sigma+2\pi)=\phi^{(n)}(\sigma)+2\pi n,\qquad n\in
\IZ_{\geq 0}.
\ee
Hence  the space of holomorphic maps  has the following decomposition
(modulo subspaces of non-zero codimension):
\be\label{decomp}
\CM(D,\IC)=\cup_{n=0}^{\infty}\CM^{(n)}(D,\IC),
\ee
where
$\CM^{(n)}(D,\IC)$ includes holomorphic maps $\varphi(z)$  such that
for the corresponding function $\phi$ the relation \eqref{monodromy}
holds. We would like to reformulate the integral \eqref{intSI} using
new variables \eqref{newpar} and the decomposition \eqref{decomp}.
Let us decompose the space of fields  $\tau(\sigma)$ on the subspace of
constant modes $\tau(\sigma)=2t$ and the orthogonal subspace of $\tau_*(\sigma)$
such that  $\int_{S^1}d\sigma\,\tau_*(\sigma)=0$.

For $\varphi\in \CM^{(n)}(D,\IC)$ the momenta \eqref{momentumOne} for
$U(1)$- and $S^1$-actions in the new variables $(\tau,\phi)$ are given
by
$$
H_{U(1)}=\frac{1}{4\pi}\int_0^{2\pi}d\sigma\,\tau(\sigma)=t,\qquad
H_{S^1}=\frac{1}{4\pi}\int_0^{2\pi}d\sigma\,\tau(\sigma)\pr_{\sigma}\phi(\sigma)=
-\frac{1}{4\pi}\int_0^{2\pi}\,d\sigma\,\pr_{\sigma}\tau(\sigma)\phi(\sigma)+
nt,
$$
where we take into account \eqref{monodromy}.
Thus  we have  the following equivalent
representation for \eqref{intSI}
\be\label{IntI}
Z(\lambda,\hbar,\mu)=\sum_{n=0}^{+\infty}\int_{\CM^{(n)}(D,\IC)}
 dt\,
 [D\tau_*]\,[D\phi]\,J(\tau_*+t)\,e^{-\frac{\mu}{4\pi}\int_{S^1}d\sigma\,
    \hbar\pr_{\sigma}\tau_*\,\phi+\mu t(\hbar n+\lambda)},
\ee where $J(\tau_*+t)$ is a Jacobain of the transformation from the
variables $(\varphi,\bar{\varphi})$ to the variables $(\tau,\phi)$.
The integration over $\phi$ leads to a delta-function with a support
on the space of  solutions of the equation \be\label{zerolocus}
\pr_{\sigma}\tau(\sigma)=0,\qquad
\tau(\sigma)=|\varphi(z)|^2|_{z=e^{\imath \sigma}}, \ee where
$\varphi(z)$ is a  holomorphic function on the disk $D$. The
solutions of \eqref{zerolocus}  are given by \be\label{critpoint}
\varphi^{(n)}(z)=\varphi_nz^n,\qquad n\in \IZ_{\geq 0} \ee and
coincide with the fixed points \eqref{fixedpt} of the $S^1$-action
on $\IP\CM(D,\IC)$. Thus the sum over  $n$ for a fixed $t$ is an
analog of the sum over $S^1$-fixed points entering
Duistermaat-Heckman formula applied to $\IP\CM(D,\IC)$. It
remains to integrate  the delta-function
$\delta(\pr_{\sigma}\tau)$ in the vicinity of each solution
\eqref{critpoint} taking into account that $\tau(\sigma)$ is a
square of a holomorphic function such that the integral
$\int_0^{2\pi}d\sigma \tau(\sigma)=2t$ is fixed. Actually we already
evaluated  this integral which is equivalent to the regularized
product \eqref{prodd} entering the Duistermaat-Heckman formula. Thus
we obtain \be\label{intone} Z(\lambda,\hbar,\mu)_{reg}=
\mu\sqrt{a\hbar}
\sum_{n=0}^{\infty}\int_{0}^{+\infty}dt\,\,\frac{(-1)^n}{(a\hbar
  /2\pi)^n n!}  e^{t\mu (\hbar n+\lambda)}=\mu \sqrt{a\hbar}\,\int_{0}^{+\infty}dt
\,e^{\mu t\lambda-\frac{2\pi}{a\hbar}e^{\mu \hbar t}}. \ee Note that
to make the integral \eqref{intone} well-define we should sum the
series for an appropriate range of the variables $\mu$ and $a$. The
integral \eqref{intone} reproduces the regularized integral
\eqref{Intone}. Taking appropriate limit \eqref{Inttwo} we recover
the expression obtained using the Duistermaat-Heckman formula. 

Using the evaluation of the integral \eqref{IntI} near the
solutions \eqref{critpoint} and summing the series one can  rewrite \eqref{IntI}
in the following form:
\be\label{IntIII}
Z(\lambda,\hbar,\mu)_{reg}=\int_0^{\infty}  dt\, \int [D\tau_*]\,
\det \Delta\,\,\delta(\Delta \tau_*)\,\delta(\pr_\sigma \tau_*|_{S^1})
e^{\mu t\lambda-\frac{2\pi}{a\hbar}e^{\mu \hbar t}},
\ee
where $\Delta$ is a Laplace operator on the disk $D$ and
now the functional integral is taken over
the space of real functions on the disk orthogonal to
the subspace of constant functions. It is easy to see that the integral over $\tau_*$
 reduces to an additional $t$-independent factor for
 $Z(\lambda,\hbar,a)_{reg}$. Combining  the variables $t$ and $\tau_*$
 into a new variable $\tau=\tau_*+t-\hbar^{-1}\ln(a\hbar/2\pi)$
 and taking the limit  $a\to +\infty$
 we obtain  the following:
\be\label{limitINT}
Z(\lambda,\hbar,\mu)=
\frac{1}{\hbar}\lim _{a \to
  \infty}\,C(a,\hbar)\,a^{-\lambda/\hbar}\,Z(\lambda,\hbar,a)_{reg}
\ee
\be
=\int [D\tau]\,\,
\det \Delta\,\,\delta(\Delta \tau)\,\delta(\pr_\sigma \tau|_{S^1})\,
e^{\frac{1}{2\pi}\int_0^{2\pi}d\sigma(\mu\lambda \tau(\sigma)
  -e^{\hbar \mu \tau(\sigma)})},
\nn\ee
where  $C(a,\hbar)$ is an  appropriate function. 
Let us note that the integral
representation \eqref{limitINT} can be directly derived
from \eqref{IntI} in the limit $a\to +\infty$. Indeed, 
in the limit $a\to \infty$ (taking into account the shift
$t\to t-\hbar^{-1}\ln(a\hbar/2\pi)$) the Jacobain becomes  field
independent and the condition on the function $\tau$ to be the square of
a holomorphic function reduces to the harmonicity condition on $\tau$
due to the expansion
$$
\Delta \ln(\tau-\hbar^{-1} \ln a\hbar/2\pi)=-\frac{\hbar}{\ln a\hbar/2\pi}\Delta
\tau_*+\cdots, \qquad  a \to +\infty.
$$
The summation over $n$ with the weight factor obtained by a proper 
integration over $n$ zeroes of $\tau$ leads to the exponential term in
\eqref{limitINT}.

To make a contact with the  representation of the
equivariant volume
integral \eqref{intSI} in terms of type B topological sigma model
described in Section 1 we note  that the
condition $\pr_{\sigma}\tau|_{S^1}=0$ imposed on  restrictions
of harmonic functions to the boundary
$S^1=\pr D$  is equivalent to
the condition $\pr_n\tau|_{S^1}=0$ where $\pr_n$ is a normal derivative to
the boundary of $D$. Therefore we have
\be\label{inttwo}
Z(\lambda,\hbar,\mu)=
\int [D\tau]\,\,
\det \Delta\,\,\delta(\Delta \tau)\,\delta(\pr_n\tau|_{S^1})\,
e^{\frac{1}{2\pi}\int_0^{2\pi}d\sigma(\mu\lambda \tau(\sigma) -e^{\mu \hbar
    \tau(\sigma)})}.
\ee
The $\delta$-functions can be replace by an integral over an
auxiliary field $\kappa(\sigma)$. Thus we obtain the following
integral representation:
\be\label{intthree}
Z(\lambda,\hbar,\mu)=
\int [D\tau]\,[D\kappa]\,
\det \Delta\,e^{\int_{D}\,\imath\,\, d\kappa\wedge *d\tau+\int_{S^1}d\sigma(
  \mu \lambda \tau(\sigma) -e^{\hbar \mu\tau(\sigma)})}\,\delta(\kappa(0)).
\ee
This functional integral is equivalent to the one entering
the formulation of the Corollary 2.1 for $\ell=0$ with
$\tau=\phi_+$ and $\kappa=\phi_-$. This can be demonstrated by
integrating over the fields $\eta$, $\theta$ and $\rho$ in the type B
model considered in previous Section.

\subsection{$T$-duality for target space  $\IC^*$}

Finally we clarify the appearance of the non-standard real structure  in the
topological type $B$-model  proposed  in Section 1 as a mirror dual to
the topological type $A$-model considered in \cite{GLO1}. To elucidate
this issue we consider a simple example of the bosonic sigma model on
$\IP^1$ with the target space $\IC^*=\IR\times S^1$.
 The mirror  symmetry in this case is straightforwardly realized
as a $T$-duality with respect to $S^1$. We will demonstrate below that
starting with a sigma model  similar to the one considered in
\cite{GLO1} we obtain after $T$-duality the topological sigma model
with the real structure on the space of  fields
considered in Section 1.

Let us given  the following  action functional
 \be\label{bosfree}
  S\,=\,\int_{\IP^1} \,\left(
  \frac{t}{2}F\wedge *F\,
  +\,F\wedge\pr\bar{\varphi}\,
  -\,F\wedge \apr\varphi\right)
\ee
\be\label{bosfreeone}
=\int_{\IP^1}\Big(\frac{t}{2}F\wedge*F\,
  -\, \imath F\wedge *d\tau\,-\, \imath F\wedge d\phi\Big)\,,
 \ee
where $\varphi=\tau+\imath \phi$ is a complex coordinate on the
cylinder $\IR\times S^1$, $\phi\sim \phi+2\pi$
 and $F=\bar{F}_zdz+F_{\zb}d\zb$ is a real valued one-form. We imply
 that  $\IP^1$ is supplied with the K\"{a}hler metric
associated with the standard K\"{a}hler form
$$
\omega=\frac{\imath}{2}\frac{dz\wedge d\zb}{(1+|z|^2)^2}.
$$
Note that (in the classical theory) the action \eqref{bosfree} does not
depend on the choice of the two-dimensional K\"{a}hler metric.
This action \eqref{bosfree} is a part of an  action of the  topological  sigma model
consider in \cite{GLO1} adopted to the case of the target space
$\IC^*$. Indeed, the integration over $F$
 gives the standard functional integral for the sigma-model
 \be
  Z\,=\,\int[DF][D\varphi]\,e^{-S}\\
  =\,\int[DF_{\zb}][D\bar{F}_z][D\varphi]\,
  \exp\Big\{-\int_{\IP^1}d^2z\left(
  t\bar{F}_zF_{\zb}\,
  -\,\imath F_{\zb}\pr_z\bar{\varphi}\,
  -\,\imath \bar{F}_z\pr_{\zb}\varphi\,
  \right)\Big\}\\
  =C(t)\,\int[D\varphi]\,\exp\Big\{- t^{-1}\int_{\IP^1}
  d^2z\,\pr_z\bar{\varphi}\,\pr_{\zb}\varphi\Big\}\,,
 \ee
where $d^2z=\imath dz\wedge d\zb$ and $C(t)$ is a function of
$t$.

The standard way to implement $T$-duality is to introduce
an auxiliary field $B=B_zdz+B_{\zb}d\zb$ and $\kappa$ and consider a
theory with the following action:
 \be\label{intonen}
  S\,=\,\i\int_{\IP^1}\,d\kappa\wedge B\,
  +\,\int_{\IP^1}\,\left(\frac{t}{2}F\wedge *F
  -\imath F \wedge *d\tau
  -\imath F\wedge B\right)\,.
 \ee
Indeed integrating over  $\kappa$ leads to a constraint $B=d\phi$
where $\phi$ is a real valued field and thus we come back to the
action \eqref{bosfreeone}.  On the other hand,  integration over $B$ leads to the  action
$$
 S=\int_{\IP^1} \,\left(
 \frac{t}{2}F\wedge *F\,
 -\,\imath F\wedge*d\tau\right),
$$
with the constraint
\be\label{Tdualcon}
 F=d\kappa.
\ee
Thus the integration over $F$ with the constraint  \eqref{Tdualcon} gives
\be\label{tact}
 S=\int_{\IP^1}\left(
 \frac{t}{2}d\kappa\wedge *d\kappa\,
 -\,\imath d\kappa\wedge*d\tau \right).
\ee
In \cite{GLO1} we consider a sigma-model without $F\wedge *F$-term
(i.e. we imply that $t=0$). Taking $t=0$ in \eqref{tact}  we obtain
 \be\label{newrealstr}
  S\,=\,-\,\i\int_{\IP^1}\, d\kappa \wedge *d\tau.
 \ee
This action is precisely the two-derivative term in
\eqref{linearTFT} where the role of $\kappa$ and $\tau$ is played by
the fields $\phi_+$ and $\phi_-$. Thus the non-standard real
structure on the fields in  \eqref{linearTFT} is a consequence of
taking a limit $t\to 0$ in the mirror dual model discussed in
\cite{GLO1}. Note that the action \eqref{newrealstr} can be
straightforwardly obtained by taking $t=0$ in \eqref{bosfreeone} and
integrating out $\phi$. Let us finally note that the action
\eqref{newrealstr} arising in the limit $t\to 0$ is analogous  to
the action functionals describing discrete light-cone quantization.
This relation will be discussed elsewhere.

\section{Conclusion}

To conclude this note  we briefly outline  some directions for future research.
The constructions of \cite{GLO1} and of this note allow several
straightforward generalizations. For instance one can consider an
equivariant  type  $A$ topological sigma model on a disk $D$
with a compact target space being (partial) flag manifolds.
Their mirror dual type $B$ topological theories are also know \cite{Gi}.
Simple examples are  provided by projective spaces $\IP^{\ell}$ and more
generally Grassmannian spaces $Gr(m,\ell+1)$. Such topological
sigma models  can be described in terms of a twisting of $\CN=2$ SUSY
gauged linear sigma models \cite{W3}, \cite{MP}. For instance in the
case of the target space $X=\IP^{\ell}$ the corresponding linear sigma
model has target space $\IC^{\ell+1}$ gauged by the diagonal action of
$U(1)$. For its mirror dual see for example  \cite{HV}. An  analog of the
correlation functions considered in \cite{GLO1} but for the target space
$\IP^{\ell}$   should be
equal to a degenerate $\mathfrak{gl}_{\ell+1}$-Whittaker function
given by
\be\label{partW}
\Psi_{\lambda_1,\ldots, \lambda_{j+1}}(x)=\int_{\CC} d\gamma\,e^{\imath \gamma x}
\prod_{j=1}^{\ell+1}\Gamma\left(\frac{\gamma-\lambda_j}{\hbar}\right).
\ee
For a detailed discussion
of the relation of \eqref{partW} to Toda chains see \cite{GLO2}.
The same expression should be equal to
an  analog of the correlation function in mirror dual
 type $B$ equivariant topological sigma
model with the target space $\IC^{\ell}$ and a superpotential
 $W(\phi)=\sum_{j=1}^{\ell}\,((\lambda_j-\lambda_{\ell+1})\phi^j-e^{\phi_j})-
e^{x-\sum_{k=1}^{\ell}\phi_k}$.
The structure of the integral \eqref{partW} is
quite transparent. The product of $\Gamma$-functions is a correlation
function in the type $A$ topological sigma model with the target space $\IC^{\ell+1}$
of the type considered in \cite{GLO1}  (as well as
a correlation function  in the mirror dual type $B$ theory)  and
the integral over $\gamma$ is a projection corresponding to an integration  over
the fields in the topological $U(1)$-gauge multiplet (over
dual scalar topological multiplet in the mirror dual
type $B$ theory). Similar reasoning can be applied to the
case of the Grassmannian target space \cite{W4}.  We will  provide a detailed discussion
of these cases in \cite{GLO3}. The case of  general partial flag
manifolds is a bit more complicated but accessible by the technique
developed in \cite{GLO4} and will be discussed elsewhere.

Let us stress that  the discussed examples of explicit
calculations of particular correlation functions in
topological theories on non-compact manifolds is not restricted to
the case of dimension two. The three- and four-dimensional examples
of such calculations have an interesting interpretation  (see e.g.
\cite{GLO1}).   These higher dimensional examples should provide
additional insights on the conjectural relation between local
Archimedean Langlands correspondence and the  mirror symmetry. We
are going to pursue these directions  elsewhere.

\vskip 1cm

\noindent {\small {\bf A.G.} {\sl Institute for Theoretical and
Experimental Physics, 117259, Moscow,  Russia; \hspace{8 cm}\,
\hphantom{xxx}  \hspace{2 mm} School of Mathematics, Trinity
College, Dublin 2, Ireland; \hspace{6 cm}\hspace{5 mm}\,
\hphantom{xxx}   \hspace{2 mm} Hamilton
Mathematics Institute, Trinity College, Dublin 2, Ireland;}}

\noindent{\small {\bf D.L.} {\sl
 Institute for Theoretical and Experimental Physics,
117259, Moscow, Russia};\\
\hphantom{xxxx} {\it E-mail address}: {\tt lebedev@itep.ru}}\\

\noindent{\small {\bf S.O.} {\sl
 Institute for Theoretical and Experimental Physics,
117259, Moscow, Russia};\\
\hphantom{xxxx} {\it E-mail address}: {\tt Sergey.Oblezin@itep.ru}}


\begin{thebibliography}{120}

\frenchspacing \smallbreak


\bibitem[ABV]{ABV} J.~Adams, D.~Barbash, and D.~A.~Vogan Jr.,
 {\it The Langlands Classification and Irreducible Characters
of Real Reductive Groups}, Progress in Mathematics, {\bf 104} Birkh\"{a}user (1992).


\bibitem[AB]{AB} M.~Atiyah, R.~Bott, {\it The momentum map and
    equivariant cohomology}, Topology {\bf 23} no. 1 (1984) pp. 1--28.

\bibitem[Au]{Au} M.~Audin, {\it Torus Actions on Symplectic
    Manifolds}, Progress in Mathematics, Birkhäuser, 2004.


\bibitem[Bu]{Bu} D.~Bump {\it Automorphic Forms and Representations},
  Cambridge  Univ. Press, Cambridge, 1998.


\bibitem[CMR]{CMR}  S.~Cordes, G.~Moore, S.~Ramgoolam,
{\it  Lectures on 2D Yang-Mills Theory, Equivariant Cohomology and
Topological Field Theories}, Proceedings of the 1994 Les Houches
school on Fluctuating Geometries, Nucl. Phys. Proc. Suppl. 41 (1995)
pp. 184--244, {\tt [hep-th/9411210]}.

\bibitem[DH]{DH} J.~J.~Duistermaat, G.~J.~Heckman,
{\it On the variation in the cohomology of the symplectic form of
the reduced phase space}, Invent. Math. {\bf69} no. 2 (1982), pp.
259--268.


\bibitem[GLO1]{GLO1} A.~Gerasimov,  D.~Lebedev,  S.~Oblezin, {\it
 Archimedean $L$-factors and Topological Field Theories I},
 {\tt [math.RT/0906.1065]}.

\bibitem[GLO2]{GLO2} A.~Gerasimov,  D.~Lebedev,  S.~Oblezin, {\it
 On q-deformed $\mathfrak{gl}_{\ell+1}$-Whittaker functions  I,II,III},
 {\tt [math.RT/0803.0145]},  {\tt [math.RT/0803.0970]},
 {\tt [math.RT/0805.3754]}.

\bibitem[GLO3]{GLO3} A.~Gerasimov,  D.~Lebedev,  S.~Oblezin, {\it
Whittaker functions and Topological Field Theories}, to appear.


\bibitem[GLO4]{GLO4} A.~Gerasimov,  D.~Lebedev,  S.~Oblezin, {\it New integral
representations of Whittaker functions for classical groups},
{\tt [math.RT/0705.2886]}.


\bibitem[Gi]{Gi} A.~Givental, {\it Stationary Phase Integrals,
Quantum Toda Lattices,
Flag Manifolds and the Mirror Conjecture}, Topics in Singularity
Theory, Amer. Math. Soc. Transl. Ser., 2 {\bf 180}, American
Mathematical Society, Providence, Rhode Island, 1997, pp. 103--115
[arXiv:alg-geom/9612001].



\bibitem[H]{H} S.~W.~Hawking, {\it Zeta function regularization of
    path integrals in curved space time}, Commun. Math. Phys. , {\bf
    55} (1977) pp. 133--148.

\bibitem[HV]{HV} K.~Hori,  C.~Vafa, {\it Mirror
    Symmetry}, {\tt [hep-th/0002222]}.

\bibitem[K]{K} M.~Kontsevich, unpublished.

\bibitem[KL]{KL} A.~Kapustin, Y.~Li, {\it
    D-Branes in Landau-Ginzburg Models and Algebraic Geometry}, JHEP
  0312:005 (2003), {\tt [hep-th/0210296]}.


\bibitem[L]{L} {\it An introduction to the Langlands program},
 Lectures presented at the Hebrew University of Jerusalem, Jerusalem,
 March 12--16, 2001. Edited by J.~Bernstein and S.~Gelbart.
Birkh\"{a}user Boston, Inc., Boston, MA, 2003.

\bibitem[MP]{MP} D. R. Morrison, M. R. Plesser, {\it
 Summing the instantons: Quantum cohomology and mirror symmetry in
 toric varieties},  Nucl. Phys. B 440 (1995) 279-354, {\tt [hep-th/9412236]}.

\bibitem[O]{O} D.~Orlov, {\it
Triangulated categories of singularities and D-branes in
Landau-Ginzburg models},  Algebraic geometry. Methods, relations,
and applications, Tr. Mat. Inst. Steklova, 246, Nauka, Moscow, 2004,
200–224;  English transl. in Proc. Steklov Inst. Math., 246
(2004), 240--262,  {\tt [math/0302304]}.


\bibitem[Se]{Se} J.~P.~Serre, {\it Facteurs locaux des fonctions
    z\^{e}ta des vari\'{e}t\'{e}s alg\'{e}braiques (d\'{e}finisions et
    conjecures)}.
 S\'{e}m. Delange-Pisot-Poitou, exp. 19, 1969/70.

\bibitem[T]{T} J.~Tate,  {\it Number theoretic Background},
in {\it Automorphic forms and $L$-functions}, Proceedings
of Symposia in Pure Mathematics, Vol. {\bf 33} (1979), part 2,
pp. 3--26.


\bibitem[Vor]{Vor} A.~Voros, {\it
Spectral Functions, Special Functions and the Selberg Zeta Function},
 Commun. Math. Phys. {\bf110}, (1987), pp. 439--465.


\bibitem[W1]{W1}\, E.~Witten, {\it Topological Quantum Field
    Theories}, Commun. Math. Phys. 117 (1988), pp. 353--386.



\bibitem[W2]{W2}\, E.~Witten, {\it Mirror Manifolds and Topological
Field Theory},  in "Essays on mirror manifolds", pp. 120--158,
Internat. Press, Hong Kong, 1992 {\tt [hep-th/9112056]}.

\bibitem[W3]{W3}\, E.~Witten, {\it
Phases of N=2 theories in two dimensions}, Nucl. Phys. B403 (1993)
159--222, {\tt [hep-th/9301042]}.

\bibitem[W4]{W4} E.~Witten, {\it
The Verlinde algebra and the cohomology of the Grassmannian},
 in Geometry, Topology, \& Physics (Cambridge, Mass., 1993),
Conf. Proc. Lecture Notes Geom. Topology 4, Internat. Press,
Cambridge, Mass., 1995,  357--422, {\tt [hep-th/9312014]}.


\end{thebibliography}
\end{document}